\documentclass{article}
\usepackage{geometry}
\usepackage[utf8]{inputenc}
\usepackage{lipsum}
\usepackage{fancyhdr}
\usepackage{xcolor}
\usepackage{soul}
\usepackage{hyperref}
\usepackage[sorting=none, maxbibnames=99]{biblatex}
\usepackage[TS1,T1]{fontenc}
\usepackage{graphicx}
\graphicspath{ {./images/} }
\usepackage{array}
\usepackage{multirow}
\usepackage{amssymb}
\usepackage{subcaption}
\newcolumntype{C}[1]{>{\centering\arraybackslash}m{#1}}
\usepackage{lineno}
\usepackage{algorithm}
\usepackage{algorithmic}
\newcommand{\argmin}{\mathop{\mathrm{argmin}}}

\title{Integrating Parcel Deliveries into a Ride-Pooling Service -\\ An Agent-Based Simulation Study}

\author{Fabian Fehn$^*$, Roman Engelhardt, Florian Dandl, Klaus Bogenberger and Fritz Busch}
\date{\textit{Chair of Traffic Engineering and Control, Technical University of Munich,\\
Arcisstr. 21, 80333 Munich, Germany\\
~$^*$corresponding author: fabian.fehn@tum.de}}

\begin{document}
\thispagestyle{empty}
\maketitle

\noindent\rule[0.5ex]{\linewidth}{1pt}
\section*{Abstract}
This paper examines the integration of freight delivery into the passenger transport of an on-demand ride-pooling service. The goal of this research is to use existing passenger trips for logistics services and thus reduce additional vehicle kilometers for freight delivery and the total number of vehicles on the road network. This is achieved by merging the need for two separate fleets into a single one by combining the services. To evaluate the potential of such a mobility-on-demand service, this paper uses an agent-based simulation framework and integrates three heuristic parcel assignment strategies into a ride-pooling fleet control algorithm. Two integration scenarios (moderate and full) are set up. While in both scenarios passengers and parcels share rides in one vehicle, in the moderate scenario no stops for parcel pick-up and delivery are allowed during a passenger ride to decrease customer inconvenience. Using real-world demand data for a case study of Munich, Germany, the two integration scenarios together with the three assignment strategies are compared to the status quo, which uses two separate vehicle fleets for passenger and logistics transport. The results indicate that the integration of logistics services into a ride-pooling service is possible and can exploit unused system capacities without deteriorating passenger transport. Depending on the assignment strategies nearly all parcels can be served until a parcel to passenger demand ratio of 1:10 while the overall fleet kilometers can be deceased compared to the status quo.
\noindent\rule[0.5ex]{\linewidth}{.1pt}
\textbf{Keywords}: integration of passengers and freight, mobility-on-demand, ride-pooling, fleet control, parcel delivery, agent-based simulation

\noindent\rule[0.5ex]{\linewidth}{1pt}

\section{Introduction}
\begin{figure}[ht]
    \centering
    \includegraphics[width=0.6\textwidth]{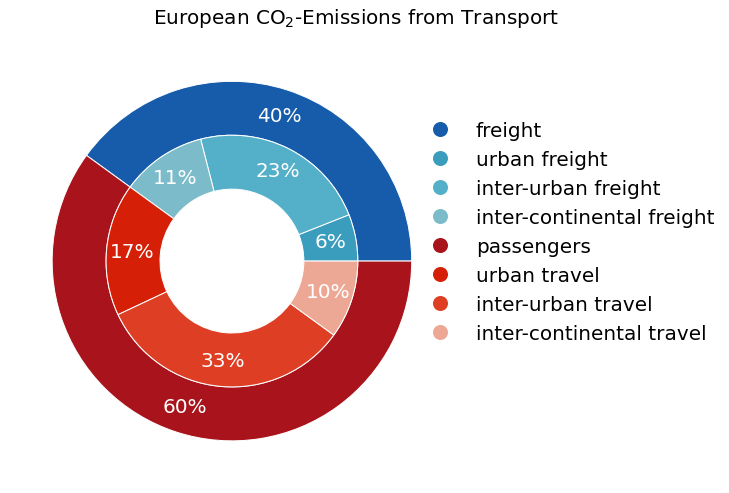}
    \caption[Caption for LOF]{Freight and passenger shares of European CO$_2$ emissions from transportation\protect\footnotemark}
    \label{fig:eu_shares}
\end{figure}

The development of urban transportation is subject to constant change. In recent years, passenger transportation has been subject to the influences of new mobility services such as car-sharing, ride-hailing, ride-pooling, or on-demand public transportation services, including autonomous shuttle buses. In addition to passenger transportation, urban freight transportation is also subject to disruptive developments. Many providers are relying on fast delivery options, such as same-day or next-day delivery options, which are increasingly provided by subcontractors. Furthermore, there has been a trend towards more environmentally friendly delivery forms, such as the use of bicycle couriers or CO$_2$-compensated delivery options in recent years. The term 'crowd logistics' is currently also heavily discussed and describes a (typically fast) shipping service outsourcing the delivery to many individuals, often private persons. The European Commission~\cite{EUCommission2013} compared the influences of passenger and freight traffic in terms of total CO$_2$ emissions in the European Union and found that passenger and freight transportation account for approximately 60\% and 40\%, respectively. More specifically, urban passenger traffic accounts for 17\%  and urban delivery traffic for 6\% of the total amount of transportation-related CO$_2$ emissions (Figure~\ref{fig:eu_shares}).

Along with that, the attitude of city residents is also evolving. While it was inconceivable a while ago that people would completely do without their privately owned cars, many citizens now rely on a combination of public transportation, sharing, and on-demand services. Generally, one can observe a more 'eco-friendly' lifestyle among consumers~\cite{PWC2021}. The so-called 'sharing economy' describes a social trend of our time, with people foregoing owning things, preferring instead to rent them or share them with others. However, there has been an increase in passenger kilometers traveled per day from 2,717 million km in 2002 over 3,080 million km in 2008 to 3,214 million km for the year 2017 in Germany~\cite{MiD2019}. Moreover, online shopping increased significantly, and accordingly, a drastic increase in home delivery traffic emerged. The annual relative growth in parcel shipments in Germany was 10.9\% in 2020 compared to 2019, with a total of 4.05 billion shipments. This is mainly due to private customer shipments, which increased by 18.6\%, whereas business customer shipments even decreased by 5.2\%~\cite{KEP2021}.

Along with the mentioned developments in transportation services and social trends, urban transportation problems have also evolved. In addition to the more transport engineering related problems, such as increased travel times or noise and pollutant emissions, transport planning problems, such as the lack of space or the division of urban districts, become also present in many European cities. Therefore, it is time to find and promote environmentally friendly and space-saving forms of transportation. In particular, the sustainable utilization of existing infrastructure plays a decisive role, as this is not directly linked to the creation of new infrastructure with the associated emissions and space consumption. With the introduction of automation in urban transportation, a time-optimized usage of the existing infrastructure could be realized, by shifting certain transportation tasks to off-peak times. Furthermore, new cost structures could establish, since the largest cost factor, the driver would be eliminated. This development could accelerate the trend towards better utilization of existing infrastructure and has therefore high disruptive potential.

One possible solution to the above-mentioned urban transportation problems, making use of current social trends and ecological and technological developments, is the integrated transport of passengers and freight in an urban context. Especially, in the field of mobility on-demand (MoD), due to its high temporal and spatial flexibility and centralized decision-making, there seem to be promising opportunities to serve passengers and freight with one fleet of vehicles. For this reason, this paper deals with strategies that enable the integrated transportation of passengers and parcels in a MoD fleet. This study evaluates the potential of such a service and assesses its impact on different stakeholders and the overall transportation system. Furthermore, this paper examines the real-world applicability and explores system boundaries with the help of a case study.

The outline of this paper is as follows: first, it describes the state of the art in research and real-world applications. Thereby, the different areas of application of integrated transport of passengers and freight, as well as existing models and solutions and case studies are discussed. The next chapter describes the methodology for the simulation and case study in Munich, Germany. Subsequently, it presents the obtained results and investigates Key Performance Indicators (KPIs) for the customers and the service operator. Last but not least, this paper discusses the findings and future research directions.

\footnotetext{According to PRIMES and TREMOVE Models of the European Commission~\cite{EUCommission2013}}

\section{State of the Art Analysis and Literature Review}
\subsection{Integrated Passenger and Freight Transportation and Real-World Applications}
The idea of transporting freight simultaneously with passengers is by no means new. As early as 1610, the first documented stagecoach traveled between Edinburgh and Leith, carrying passengers as well as smaller parcels on its journey~\cite{Lay1992}. Nowadays, the joint transport of passengers and parcels is also not uncommon. Today's passenger aircraft typically also handle freight. The combined transport of people and goods is also quite common in maritime shipping and ferry service. However, much fewer services exist for combined transportation in an urban context. Yet, intelligent solutions are needed here in particular, as urban transport phenomena, such as increased travel times~\cite{Inrix2021}, space shortages, noise emissions, and poor air quality have worsened considerably in recent years, according to the German Ministry of Environment~\cite{BMUV2022}. For this reason, a couple of smart approaches have already emerged that could lead to an improvement in the aforementioned combined transportation of passengers and freight in an urban context. To classify the existing urban concepts and to investigate existing simulation studies, this paper analyses the existing concepts of integrated passenger and freight transportation. A classification between the different forms can be made by distinguishing between public transportation (rail- and road-based) and individual transportation.

The integration of rail-based public transportation systems and freight transportation can have many faces, thereby one of the most researched forms is the railroad transportation sector~\cite{Li2021, Larodde2020, Behiri2018}. The combined transportation in urban subway systems, especially in off-peak times, is also a very promising approach and has therefore already been examined in several research papers~\cite{He2019, Zhou2019, Dampier2015}. Combinations of Personal Rapid Transit and Freight Rapid Transit are also subject of research and combine the two transportation streams on rail~\cite{Cochrane2016, Fatnassi2014}. Another interesting approach is to use urban tramways~\cite{Delanghe2019}. The city of Zurich launched an operational service called Cargo-Tram and E-Tram in 2003~\cite{Zurich2022}, which acts as a recycling center on rails and transports large and heavy waste and electronic devices in off-peak times, however, without passengers on board. The Karlsruhe Institue of Technology and the University Frankfurt am Main of Applied Sciences research integrated passenger and freight transportation approaches in the 'LogIKTram' and 'LastMileTram' research projects~\cite{Kagerbauer2022, Schocke2020}. Another solution could be the combined transport of passengers and freight in urban cable cars~\cite{Pernkopf2021}.
Road-based public transportation systems are also a matter of discussion when it comes to integrating passenger and freight transport. Especially, bus systems are subject of current research~\cite{Bruzzone2021, Ghilas2016, Arvidsson2016, Trentini2012}. An interesting concept study in this category is the 'Freight*Bus', which according to the persons in charge could change the economic and environmental costs of passenger and freight transportation in modern cities~\cite{Frost2008}. One of the probably best-known examples of a road-based, multi-purpose vehicle approach is the Toyota e-Palette concept, which was presented in the context of the Olympic Games 2020 in Tokyo~\cite{Toyota2018}. The e-Palette concept aims at combining multiple purposes, such as passenger and freight transportation or mobile shopping facilities, on one autonomous vehicle platform.

Apart from public transportation infrastructure, there is also the possibility to integrate passenger and freight flows into private transportation. In the case of crowdsourced delivery, the introduction of new vehicles or infrastructure becomes superfluous, because these approaches rely on already existing vehicles or passenger trips. The concept is a kind of ride-sharing service for parcels, where the parcels are picked up and driven to their destination by registered private individuals for a small fee~\cite{Alnaggar2021, Sampaio2019} or bonus points~\cite{Liu2019}. Privately operated transportation services, such as taxi services, should also be mentioned in this context and are themselves already subject of research~\cite{Chen2015, Li2014_1}. Especially, centrally controlled MoD vehicle fleets reveal a high potential for the combined transportation of passengers and freight, as the decision-making (i.e. vehicle assignment and routing) is bundled. In autonomous mobility on-demand (AMoD) systems, the driver is no longer a cost- and time-limiting factor, thus enabling new application scenarios for combined transportation~\cite{Schlenther2020}. Apart from the car as a transport vehicle, the combined transportation of passengers and freight on two-wheeled vehicles (i.e. bicycles and motorcycles) has also already been studied~\cite{Howe2013}.

The combination of MoD for passengers and parcels is the core research field of this paper and this research refers to it as ride-parcel-pooling (RPP), in reference to the widely used term ride-pooling, which describes the joint and simultaneous transportation of passengers with similar origin-destination relationships in one vehicle.

\subsection{Integrated Transport in Mobility-on-Demand Research}

Several studies dealt with the control and the efficiency of mobility on-demand services in recent years. Ride-hailing services can reduce the overall needed fleet size compared to private vehicles~\cite{Fagnant.2015b} or car-sharing services~\cite{Dandl.2019b} because of higher temporal utilization, even without shared rides. To also reduce vehicle kilometers in a mobility system, rides have to be shared using ride-pooling services to overcome empty pick-up trips~\cite{Engelhardt2019, Ruch.2020}. Nevertheless, it has been shown that the efficiency of pooling, i.e. the probability of finding shareable trips, heavily depends on fleet size and especially the overall demand ~\cite{Tachet.2017,Bilali.2020}. While the utilization of fleet vehicles can be increased by pro-actively distributing idle vehicles (i.e. re-balancing) according to expected demand ~\cite{Dandl.2019,Syed.2021}, vehicles remain idle when no passenger demand is present. The integration of parcel demand can fill these idle items, however, new fleet control algorithms have to be developed. Most state-of-the-art ride-pooling assignment algorithms heavily utilize explicit time constraints on customer pick-ups and in-vehicle travel times enabling graph-based approaches~\cite{Santi.2014,AlonsoMora2017,Simonetto.2019}. Nevertheless, these approaches become computationally intractable if these time constraints are relaxed, which is the case for most parcel assignment problems.

The integration of on-demand passenger and freight transportation is a heavily discussed topic in the expert community. The underlying optimization problem for the routing of vehicle fleets is a variant of the Vehicle Routing Problem (VRP). The VRP is one of the most studied optimization problems in transportation research and has already been addressed in numerous publications. A very good overview of the different approaches can be found in the literature reviews by Eksioglu et al.~\cite{Eksioglu2009} for the general VRP and Kim et al.~\cite{Kim2015} for the city VRP. In terms of joint transport of passengers and freight, Cavallero and Nocera~\cite{Cavallaro2021} conducted a concept-centric literature review to classify the different integration concepts.

The present paper focuses on the integration of MoD of passengers and same-day delivery of freight and classifies the existing literature on the subject according to the categories: overall integration concept, passenger mobility, and logistics service characteristics, optimization approach, and, last but not least, the real-world examination of the results in a case study (Table~\ref{table:literature}).\newline

This chapter collects the different existing approaches, case studies, and mathematical models and algorithms dealing with the modeling of integrated passenger and freight transport in MoD ride-pooling and ride-hailing services. Mourad et al.~\cite{Mourad2019_1} summarize models and algorithms for optimizing shared mobility, which serves as an input for the literature collection in this chapter and Table~\ref{table:literature}. In their paper the authors summarize the recent research activities in the field, including different optimization approaches, to provide guidelines and give promising directions for future research. In the following, this paper gives a brief overview of the methodology and the results of existing research, and the respective contribution to the scientific discourse.

\begin{table}[ht!]
\centering
\begin{tabular}{C{0,7cm}|C{2,4cm}|C{1,5cm}|C{1,2cm}|C{1,2cm}|C{1,2cm}|C{1,2cm}|C{1,0cm}|C{1,0cm}|C{1,1cm}}
\multirow{2}{*}{\parbox{1cm}{Ref.}} & \multirow{2}{*}{\parbox{1cm}{\centering{Author, Year}}} & \multirow{2}{*}{Mode} & \multicolumn{2}{c|}{MoD Service} & \multicolumn{2}{c|}{Logistics Service} & \multicolumn{2}{c|}{Optimization} & \multirow{2}{*}{\parbox{1cm}{\centering{Case Study}}}\\
&  &  & hailing & pooling & imme-diate & sche-duled & static & dyna-mic & \\ [0.5ex] 
\hline\hline
\cite{Li2014_1} & B. Li et al., 2014 & Taxi & \checkmark & - & \checkmark & - & \checkmark & \checkmark & \checkmark \\
\cite{Li2014_2} & L. Li et al., 2014 & Multi & - & - & \checkmark & - & - & \checkmark & -  \\
\cite{Ngoc-Quang2015} & Ngoc-Quang et al., 2015 & Taxi & \checkmark & - & \checkmark & - & - & \checkmark & \checkmark \\
\cite{Chen2015} & C. Chen et al., 2015 & Taxi & \checkmark & - & \checkmark & - & - & \checkmark & - \\
\cite{Ronald2016} & Ronald et al., 2016 & MoD & \checkmark & - & \checkmark & - & - & \checkmark & \checkmark \\
\cite{SotoSetzke2017} & Soto Setzke et al., 2017 & MoD & \checkmark & - & - & \checkmark & - & \checkmark & \checkmark \\
\cite{Chen2017} & C. Chen et al., 2017 & Taxi & \checkmark & - & \checkmark & - & \checkmark & \checkmark & \checkmark \\
\cite{Kafle2017} & Kafle et al., 2017 & Ped., Cycl. & - & - & \checkmark & - & - & \checkmark & - \\
\cite{Beirigo2018} & Beirigo et al., 2018 & AMoD & - & \checkmark & \checkmark & - & - & \checkmark & - \\
\cite{Qi2018} & Qi et al.,\newline 2018 & Car & - & - & \checkmark & - & \checkmark & - & - \\
\cite{Wang2018} & Wang et al., 2018 & Car & - & - & \checkmark & - & - & \checkmark & \checkmark \\
\cite{Arslan2018} & Arslan et al., 2018 & Car & \checkmark & - & \checkmark & - & - & \checkmark & - \\
\cite{Mourad2019_2} & Mourad et al., 2019 & Multi & - & \checkmark & \checkmark & - & - & \checkmark & - \\
\cite{Najafabadi2019} & Najaf Abadi, 2019 & Taxi & \checkmark & - & - & \checkmark & - & \checkmark & \checkmark \\
\cite{Chen2020} & Y. Chen et al., 2020 & Taxi & \checkmark & - & \checkmark & - & - & \checkmark & \checkmark \\
\cite{Manchella2020} & Manchella et al., 2020 & Taxi & - & \checkmark & \checkmark & - & - & \checkmark & \checkmark \\
\cite{Schlenther2020} & Schlenther et al., 2020 & MoD & - & \checkmark & \checkmark & - & - & \checkmark & \checkmark \\
\cite{VanTholen2021} & Van der Tholen et al., 2021 & AMoD & - & \checkmark & \checkmark & - & - & \checkmark & - \\ 
\cite{Alho2021} & Alho et al., 2021 & MoD & \checkmark & \checkmark & - & \checkmark & - & \checkmark & \checkmark \\
\cite{Fehn2021} & Fehn et al., 2021 & MoD & - & \checkmark & - & \checkmark & \checkmark & - & \checkmark \\
\cite{Meinhardt2022} & Meinhardt et al., 2022 & MoD & - & \checkmark & \checkmark & - & - & \checkmark & \checkmark \\
\cite{Zhang2022} & Zhang et al., 2022 & AMoD & - & \checkmark & \checkmark & - & - & \checkmark & \checkmark \\
\end{tabular}
\caption{Existing Models in Literature and assorted research characteristics (i.e. Transport Mode, Type of MoD Service, Optimization Approach, and Case Study)}
\label{table:literature}
\end{table}

B. Li et al.~\cite{Li2014_2} investigate a Share-a-Ride Problem (SARP) for passengers and parcels sharing taxis. Therefore, the authors propose a reduced SARP and use the Freight Insertion Problem (FIP) to insert parcel requests into the vehicle routes. They evaluate their model with historical taxi data from San Francisco and performed a numerical study of static and dynamic scenarios. L. Li et al.~\cite{Li2014_1} research a multi-agent cooperative inter-modal freight transportation planning approach for multiple inter-modal freight transport operators. In a simulation study, they show the potential of their cooperative planning approach. Ngoc-Quang et al.~\cite{Ngoc-Quang2015} introduce a practical hybrid transportation model for serving passengers and parcels in the same fleet of taxis. They show the feasibility and efficacy of their time-dependent model using two different heuristic algorithms. In the next step, the authors use a real-world data set of the 'Tokyo-Musen' taxi company to simulate the joint transport of passengers and freight. Therefore, they introduce a practical hybrid transportation model for Tokyo city to handle passengers and parcels in the same vehicle. The paper adopts the model of Li et al.~\cite{Li2014_1} to a real-life use case. The simulation results are subsequently analyzed on the total benefit for the taxi provider, the accumulated fleet distance over the day, and the number of used taxis and served requests. Chen and Pan~\cite{Chen2015} propose a taxi fleet-based crowd-sourcing solution to last-mile delivery for an e-commerce delivery use case. Their approach relies on a two-phase decision model, first an offline taxi trajectory mining, and second an online package for routing and taxi scheduling. Ronald et al.~\cite{Ronald2016} create a simulation of a VRP and a FIP. They explore the integrated MoD passenger and freight transportation problem using a simulation of an on-demand transportation scheme. They investigate three different scenarios: 1) each shop has its delivery vehicles, 2) all shops share delivery vehicles, and 3) a co-modal scheme where passengers and parcels can share vehicles. The authors show, that their co-modality approach, using household survey data from Melbourne to generate demand for passengers and parcel customers, can provide an improved experience for both, operators and customers and that on-demand co-modality is more resilient to uneven or unexpected demands, and provides more options for travel compared to the other two scenarios. Soto Setzke et al. \cite{SotoSetzke2017} suggest an algorithm optimizing the assignment of drivers to transport requests, using transport routes and time constraints as inputs. They evaluate their algorithm based on a mobility data set from a major German city. Chen et al.~\cite{Chen2017} try to exploit taxi trips by creating relays of multiple trips to integrate the transport of parcels, without degrading the quality of passenger services. They develop a city-wide parcel delivery system, leveraging a crowd of taxis in Hangzhou, China. The authors introduce a two-phase framework, which in the first phase mines historical taxi data offline and in the second phase uses an online adaptive taxi scheduling algorithm to iteratively find near-optimal delivery paths. Their results show, that over 85\% of parcels can be delivered within 8 hours, with around 4.2 trans-shipments/relays on average. Kafle et al.~\cite{Kafle2017} consider cyclists and pedestrians as crowdsource for last- and first-mile operations. The distributors request deliveries by submitting bids to the logistics operator. The authors find, that the total truck miles traveled and the total cost can be reduced compared to pure-truck delivery.
Beirigo et al.~\cite{Beirigo2018} model a variation of the People and Freight Integrated Transport (PFIT) Problem. They find that mixed-purpose fleets perform on average 11\% better than single-purpose fleets. Qi et al.~\cite{Qi2018} present new logistics planning models and managerial insights. The proposed model uses open-loop car routes to assign passenger vehicles to parcels and fulfill the last-mile delivery job. Their findings suggest that crowd-sourcing shared mobility is not as scalable as conventional truck-based logistics systems in terms of operating cost. However, the results also show that reducing the truck fleet size and exploiting additional operational freedom, like avoiding high-demand areas and peak hours could be of interest. Wang et al.~\cite{Wang2018} use a crowd of connected vehicles to pick-up and deliver parcels, by connecting them with a logistics provider. In their simulation study, they used real-world car trips and assigned the parcel trips accordingly. The authors believe that ride-sharing will be a core service for connected vehicles, which they refer to as ride-sharing as a service. Arslan et al. \cite{Arslan2018} consider a service platform for crowdsourced delivery using existing journeys. The authors propose a rolling horizon framework and searched for an exact solution to the matching problem. Their results suggest that ad-hoc drivers have the potential to save up to 37\% of vehicle kilometers compared to traditional delivery systems. Mourad~\cite{Mourad2019_2} investigates in his PhD thesis how to synchronize people and freight flows. He develops a matching algorithm using an Adaptive Large Neighborhood Search (ALNS) heuristic and tests the approach in several experiments. Najaf Abadi~\cite{Najafabadi2019} develops an on-demand dynamic crowd-shipping system and tried to take advantage of the unused capacities in taxis. In her PhD-thesis she investigates the effects of an on-demand dynamic crowd-shipping system, using the publicly available New York City taxi data and freight demand from structural data. She studies the effects on trip cost, vehicle miles traveled, and peoples' travel behavior and finds that the proposed crowd-shipping model has a substantial positive impact on the average total system-wide vehicle miles savings for all scenarios, ranging from 47\% to 50\%. Chen et al.~\cite{Chen2020} present a new parcel delivery scheme that takes advantage of multi-hop ride-sharing. They tackle the assignment problem using a two-phase solution, which first predicts the passenger orders and then plans the parcel delivery routes using real-world data from the city of Chengdu. Their results suggest, that the successful delivery rate may reach 95\% on average daytime and is at most 46.9\% higher than those of the benchmarks. Manchella et al. \cite{Manchella2020} develop a deep reinforcement learning algorithm for joint passengers and goods transportation using the publicly available New York City taxi data. The proposed algorithm pools passengers and delivers goods using a multi-hop transit method. The multi-agent simulations carried out in the paper show, that the approach achieves 30\% higher fleet utilization and 35\% higher fuel efficiency in comparison to 1) model-free approaches where vehicles transport a combination of passengers and goods without the use of multi-hop transit, and 2) model-free approaches where vehicles exclusively transport either passengers or goods. Schlenther et al.~\cite{Schlenther2020} propose a methodology to simulate the behavior of a vehicle fleet serving passengers and freight within an urban traffic system and evaluate its performance using the simulation environment MATSim. Their results suggest, that the vehicle miles traveled for freight purposes increase due to additional access and egress trips. Van der Tholen et al.~\cite{VanTholen2021} present a method to estimate the optimal capacity for the passenger and parcel compartments of AMoD systems. They aim at creating an optimal routing schedule between a randomly generated set of pick-up and drop-off requests of passengers and parcels. Alho et al.~\cite{Alho2021} investigate the use of MoD services performing same-day parcel deliveries. Therefore, they evaluate a cargo-hitchhiking service for e-commerce using MoD vehicles in a simulation-based approach. Their research aims at testing MoD-based solutions using an agent- and activity-based simulation platform for the joint transport of passengers and freight. The authors obtain e-commerce demand carrier data collected in Singapore and investigate operational scenarios fulfilling the deliveries with MoD service vehicles on existing passenger flows. The results of a case study in Singapore indicate that the MoD services have the potential to fulfill a considerable amount of parcel deliveries and decrease freight vehicle traffic and total vehicle kilometers traveled without compromising the quality of MoD for passenger travel. Fehn et al.~\cite{Fehn2021} investigate the combined transport of passengers and parcels using historical MoD trips and parcel demand. The authors set up a static optimization problem for the combined transport of MoD passengers and intra-city parcel shipments. Therefore, the paper uses real-world individual transportation and parcel data from the city of Munich and matches the parcel data in a static optimization approach on pre-computed MoD passenger trips. The results show, that depending on the simulated scenario about 80\% of the distance traveled to provide the logistics services could be saved, offering a 100\% delivery rate. The study also gives insights into the positive environmental impacts of such an integrated transportation approach. Meinhardt et al.~\cite{Meinhardt2022} study the joint transport of passengers and freight using the Multi-Agent transport Simulation (MATSim) software. The authors apply real-world data from Berlin in the simulation and showed, that when assuming a relatively large vehicle fleet, passenger waiting time statistics barely change. Furthermore, a rough cost analysis suggests a large saving potential when using AMoD vehicles instead of a privately owned vehicle fleet. Zhang et al.~\cite{Zhang2022} propose an AMoD system performing joint ride-sharing and parcel delivery. The distributed approach, setting up a mixed-integer linear programming problem and solving it with a Lagrangian dual decomposition method, shows near-optimal solutions in reduced computation time. Nevertheless, due to the complexity of the problem, only 10 passenger and 5 parcel requests could be modeled in reasonable computational time.

\subsection{Research Gap and Contribution}
As the literature review reveals, the range of possible combinations of passengers and freight is very wide. It comprises different transportation modes, like rail-based public transportation services or individual crowdsourced approaches to the idea of central assignment of passengers and parcels to a vehicle fleet, like in the case of MoD transportation services. In this context, the solution approaches in the MoD area emerge as very promising due to their high flexibility in terms of spatio-temporal network coverage and relatively low capacity utilization when no passenger demand is present. 

Most recent studies focused either on temporally separated integration of passenger and freight or the integration into ride-hailing service, where passengers do not share the rides. Furthermore, they assume time windows for the pick-up and drop-off of parcels, which simplifies the assignment, but does not exploit the full integration potential of existing passenger trips. This study focuses on the integration of freight transport into the operation of an MoD ride-pooling service. Thereby, special focus is put on the simultaneous transport of passengers and freight (i.e. passengers and parcels can be on board the same vehicle). A dynamic simulation environment is proposed where decisions to serve customers and or parcels have to be made online to model a realistic setting. As can be seen in Table~\ref{table:literature}, the authors are only aware of the approach of Alho et al.~\cite{Alho2021} to meet similar contributions. Nevertheless, a different approach to integration is investigated in this study. While recent studies focus on exploiting idle times of MoD fleets explicitly for logistic services and/or modeled the logistics service as an as-soon-as-possible delivery service (i.e. employing strict time constraints on parcel pick-up and delivery), the goal of this study is to actively integrate parcel pick-up and drop-off into vehicle routes. No explicit time constraints are enforced on parcel pick-up and delivery, but the goal is the insertion into vehicle schedules resulting from the underlying MoD service to minimize the additional driven fleet kilometers for the integrated logistic service. To solve the assignment problem of which parcel to assign when into the dynamically evolving vehicle schedules, three heuristic assignment approaches are proposed. The efficiency of this integration and the developed assignment approaches is evaluated regarding the impacts on operator, customers and traffic effects, based on real-world logistics demand, within a case study for Munich, Germany.

\section{Methodology}

The methodology of this study is described in the following three sections: 1) the different scenarios examined in this paper to integrate the parcel delivery into the MoD service are defined, 2) the simulation framework and the proposed algorithms to control the fleet of vehicles to serve passenger, as well as parcel demand, is introduced and 3) the simulation inputs based on real-world logistics and travel demand data for a case study in Munich, Germany is described.

\subsection{Service Definition and Scenarios of Integration}

\begin{figure}[!b]
    \centering
    \includegraphics[width=0.32\textwidth]{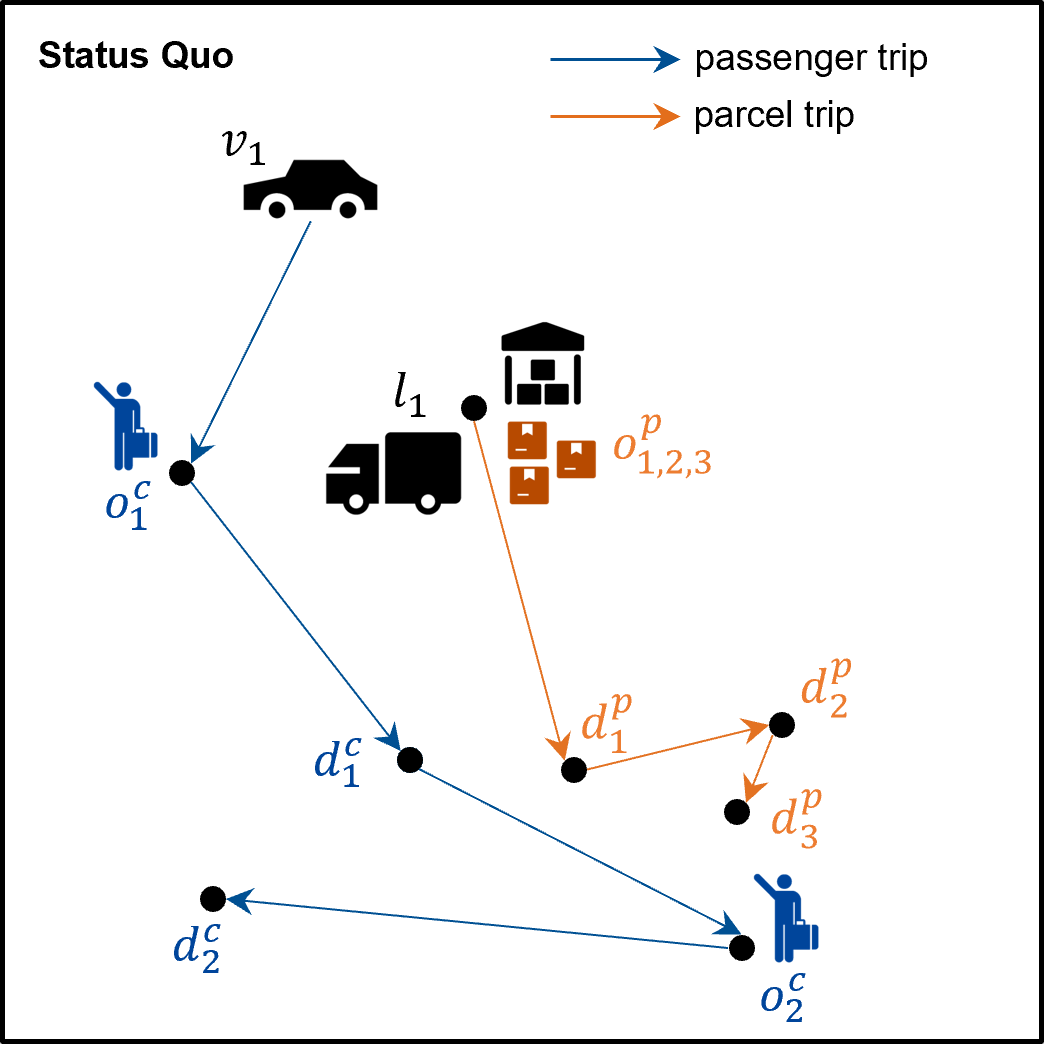}
    \includegraphics[width=0.32\textwidth]{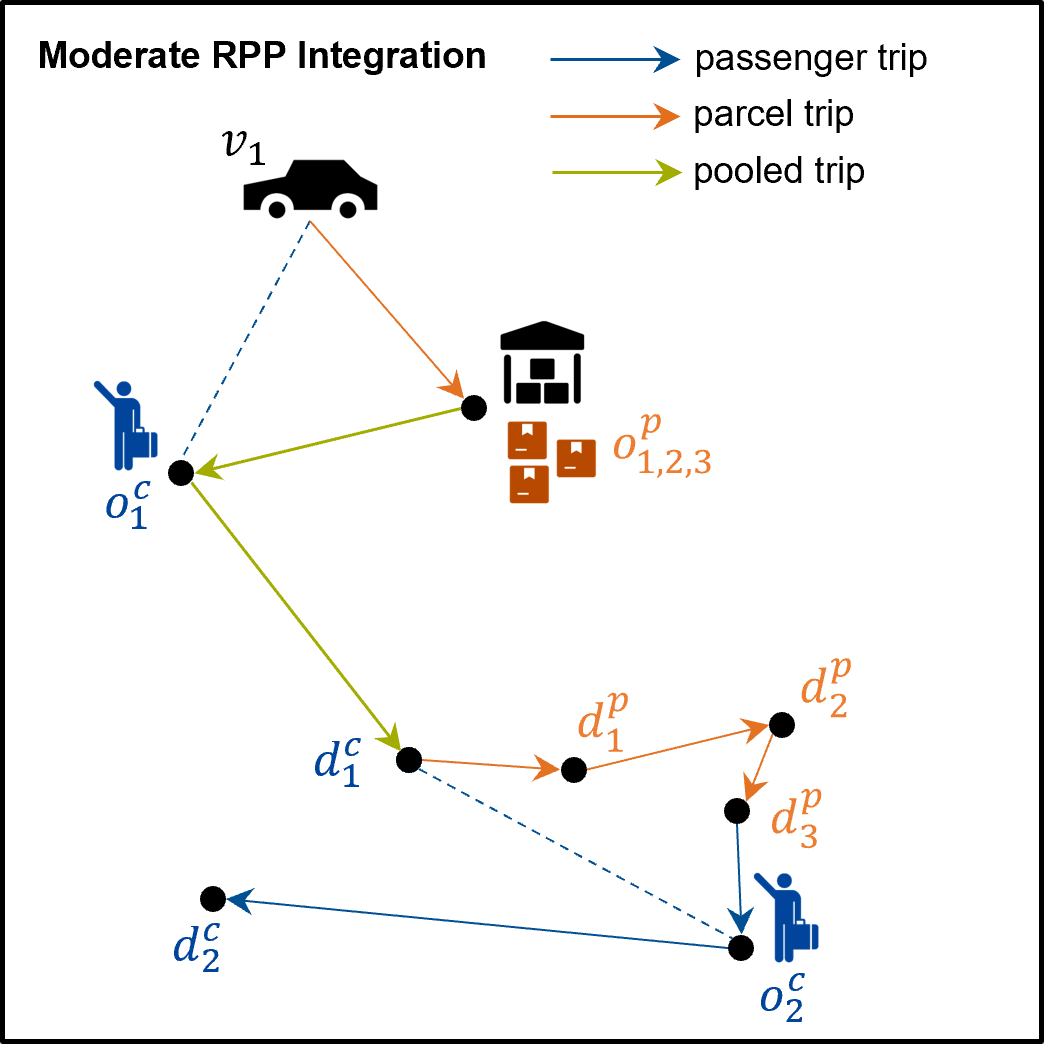}
    \includegraphics[width=0.32\textwidth]{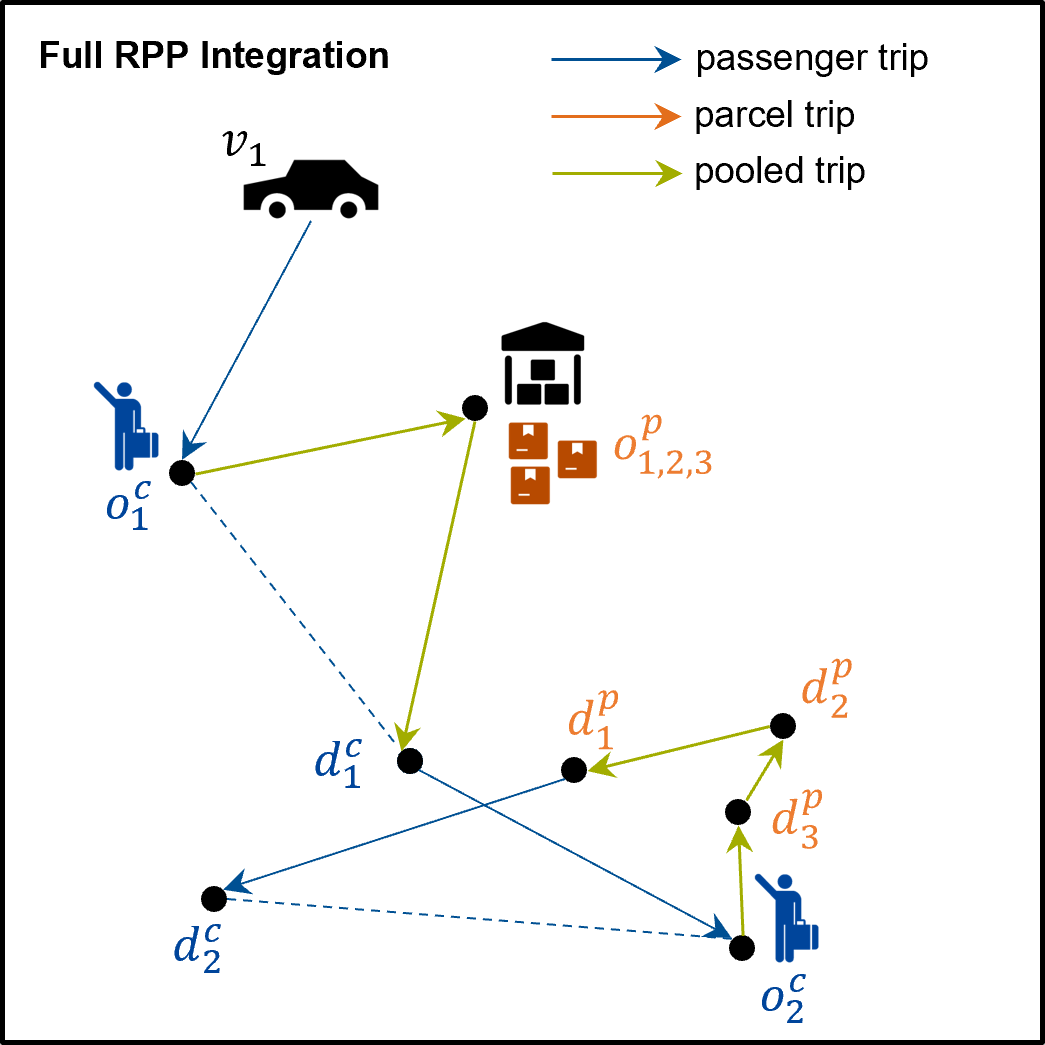}
    \caption{Illustration of the RPP Scenarios with Urban Logistics Depot}
    \label{fig:with_depot}
\end{figure}

This research considers a Ride-Parcel-Pooling (RPP) service to investigate the integrated transport of passengers and freight within the same MoD vehicle fleet. The service consists of a MoD ride-pooling vehicle fleet and a central operator for decision-making (i.e. vehicle assignment and routing). In the envisioned RPP service, the transport of passengers has priority over parcel transport requests, as the service for passengers should not deteriorate. The service assumes that the parcels are transported from urban logistics hubs to the customers. It is assumed that a pick-up or drop-off of the parcels at the customer's premises is possible at any time, which in reality could be realized by small parcel lockers. Therefore, no explicit business hours are modeled for the logistics service. Nevertheless, parcel pick-ups and deliveries are integrated into passenger trips, therefore most parcel pick-ups and deliveries will occur during daytime when most passenger demand is present.  The characteristics of the parcels in size and weight are assumed to be boxes, which can be carried by one person and fit into the trunk of a conventional passenger vehicle. The driving comfort of the passengers does not deteriorate while parcels are on board the vehicle, as it is assumed that parcels are transported spatially separated, in the trunks of the vehicles. Furthermore, the service performs a typical logistics service and does not guarantee a certain delivery period during the day. The goal of the operator is to provide an additional service without compromising too much on the additional vehicle kilometers covered by the fleet. It also allows the fleet operator to achieve higher utilization and occupancy rates for the vehicle fleet, and to keep the transport service running even during periods of low passenger demand, which in turn could lead to higher customer satisfaction. Additionally, the RPP service could contribute to the goal of the so-called 'livable city', as two vehicle fleets (i.e. logistics fleet and ODM fleet) become one and travel kilometers can be saved by pooling passengers and parcels.

To be able to evaluate the RPP service, three scenarios are introduced. Scenario No. 1 displays the current \textit{Status Quo}, where passenger and parcel requests are served by two independent vehicle fleets, one specialized in a ride-pooling MoD service, and the other on a typical urban pick-up and delivery logistics use case. In Scenario No. 2, denoted by \textit{Moderate RPP Integration}, the logistics service is integrated into the MoD service and both, passenger and parcel requests, are served by the same MoD vehicle fleet. However, there is the requirement, that no parcels are collected or delivered during a passenger trip so that passengers do not experience the parcel pick-up and drop-off (PUDO) or take detours due to the additional parcel transport. Scenario No. 3, \textit{Full RPP Integration} loosens this requirement and thus allows the collection or delivery of a parcel during a passenger trip. In this scenario, a passenger's journey may be extended or a detour may have to be accepted due to the additional parcel transport. Figure~\ref{fig:with_depot} illustrates the respective simulation scenarios and gives a schematic overview of potential assignments of requests to vehicles and the resulting routes. It is important to note that a passenger request always consists of an origin-destination (OD) pair and that a parcel request always starts at a parcel depot. Thus, for parcel transport, either origin or destination consists of a depot location and the respective other point corresponds to the pickup or delivery location of a logistics end customer.

\subsection{Agent-based Simulation and Fleet Control Strategies}

To model the integration of logistics in an on-demand ride-pooling service, the agent-based simulation framework 'FleetPy'~\cite{FleetPy} is extended for this use case. The framework consists of four main agents: 1) customers requesting trips from the fleet operator; 2) parcels that need to be transported by the service; 3) a fleet operator offering the service and assigning schedules for its fleet of vehicles to pick-up and drop-off customers and parcels, and 4) vehicles traveling along these assigned schedules within a street network and fulfilling the corresponding pick-up and drop-off tasks. All symbols used in the following can be found in Table~\ref{table:symbols}.

\begin{table}[!t]
\centering
\begin{tabular}{l l}
 \hline
 \textbf{Symbol} & \textbf{Description} \\
 \hline
 $G$ & Street network, consisting of nodes ($N$) and edges ($E$) \\
 $N$ & Network nodes \\
 $E$ & Network edges \\
 $r_i^c$ & Request of customer $i$ \\
 $o_i^c$, $d_i^c$, $t_i^c$  & Origin, destination, and request time of MoD customer $i$ \\
 $r_i^p$ & Request of parcel $i$ \\
 $o_i^p$, $d_i^p$ & Origin, and destination of parcel $i$ \\
 $v$ & MoD vehicle\\
 $V$ & MoD vehicle fleet \\
 $c_v^c$,  $c_v^p$ & Vehicle capacity for customers, and parcels \\
 $t_{max}^{wait}$, $t_{max}^{travel}$ & Maximum waiting, and travel time for a customer \\
 ${\Delta}$ & Detour factor \\
 $tt_{i}^{direct}$ & Direct travel time for request $i$ \\
 $\psi_k$ & Feasible vehicle schedule \\
 $R_\psi$, $P_\psi$ & Set of customer and parcel requests \\
 $t_b$ & Time needed for boarding and alighting \\
 $\phi(\psi_k)$ & Objective function for the rating of vehicle schedule $\psi_k$ \\
 $d(\psi_k)$ & Distance to drive to complete vehicle schedule $\psi_k$ \\
 $P$ & Assignment reward to prioritize passenger transport \\
 $T_{repo}$ & Period of MoD fleet re-balancing \\
 $T_{p}^{max}$ & Time when remaining parcels in the vehicles are delivered \\
 $\tau_{th}$ & Parcel detour threshold parameter for assignment \\
 \hline
\end{tabular}
\caption{List of symbols for the methodology section}
\label{table:symbols}
\end{table}

The street network $G=(N,E)$ consists of nodes $N$ and edges $E$ connecting these nodes. Each edge is associated with a travel time and a distance. A customer request $r_i^c = (o_i^c, d_i^c, t_i^c)$ is represented by its origin location $o_i^c \in N$, the destination location $d_i^c \in N$ and the request time $t_i^c$. In this study, it is assumed that the parcel delivery request was submitted at least the day before and is therefore known in advance for the whole simulation period. Thus, a parcel request $r_i^p = (o_i^p, d_i^p)$ is only represented by the corresponding origin (pick-up) location $o_i^p$ and destination (drop-off) location $d_i^p$. 

The goal of the fleet operator is to assign schedules to its vehicles $v \in V$ to serve customer and parcel requests. A schedule $\psi$ describes the sequence of pick-up and drop-off stops assigned to a vehicle. A schedule is considered feasible, if:
\begin{itemize}
    \item the drop-off succeeds the pick-up for each request.
    \item at no point during the schedule the maximum passenger capacity $c_v^c$ and maximum parcel capacity $c_v^p$ is exceeded by on board passengers and parcels, respectively.
    \item for each customer request $r_i^c$, the waiting time (time between request time $t_i^c$ and expected pick-up) does not exceed $t_{max}^{wait}$.
    \item the in-vehicle time of each customer $r_i^c$ does not exceed $t_{max}^{travel} = (1+\Delta) tt_{i}^{direct}$, with the direct travel time from its origin to destination $tt_{i}^{direct}$ and a detour factor $\Delta$.
\end{itemize}
In case of modelling the \textit{Moderate RPP Integration}, an additional constraint is added for a schedule to be feasible:
\begin{itemize}
    \item while passengers are on board the vehicle, no stop is allowed where only parcels are picked up or dropped off; the transport of parcels and a stop for other passengers, however, is possible.
\end{itemize}

A feasible vehicle schedule $\psi_k(v;R_\psi, P_\psi)$ is defined as the $k$-th feasible permutation of stops of vehicle $v$ to serve the set of customer requests $R_\psi$ and parcel requests $P_\psi$ within the schedule. Stops are associated with the location, boarding and alighting customers or parcels and the time needed for boarding or alighting $t_b$. In between stops, vehicles travel along the fastest route in the network $G$.

Schedules are rated by an objective function $\phi(\psi_k(v;R_\psi, P_\psi))$. The goal of the fleet operator is to assign schedules minimizing the aggregated objective function for all of its vehicles. In this study, the objective function is defined as:
\begin{equation}
    \label{eq:obj}
    \phi(\psi_k(v;R_\psi, P_\psi)) = d( \psi_k(v;R_\psi, P_\psi) ) - P (|R_\psi| + |P_\psi|) .
\end{equation}
$d( \psi_k(v;R_\psi, P_\psi) )$ refers to the distance to drive to complete the schedule. $|R_\psi|$ and $|P_\psi|$ are the number of customers and parcels to be served with the schedule, respectively. $P$ is a large assignment reward to prioritize serving customers and parcels over minimizing the driven distance.

\begin{figure}[ht]
    \centering
    \includegraphics[width=0.8\textwidth]{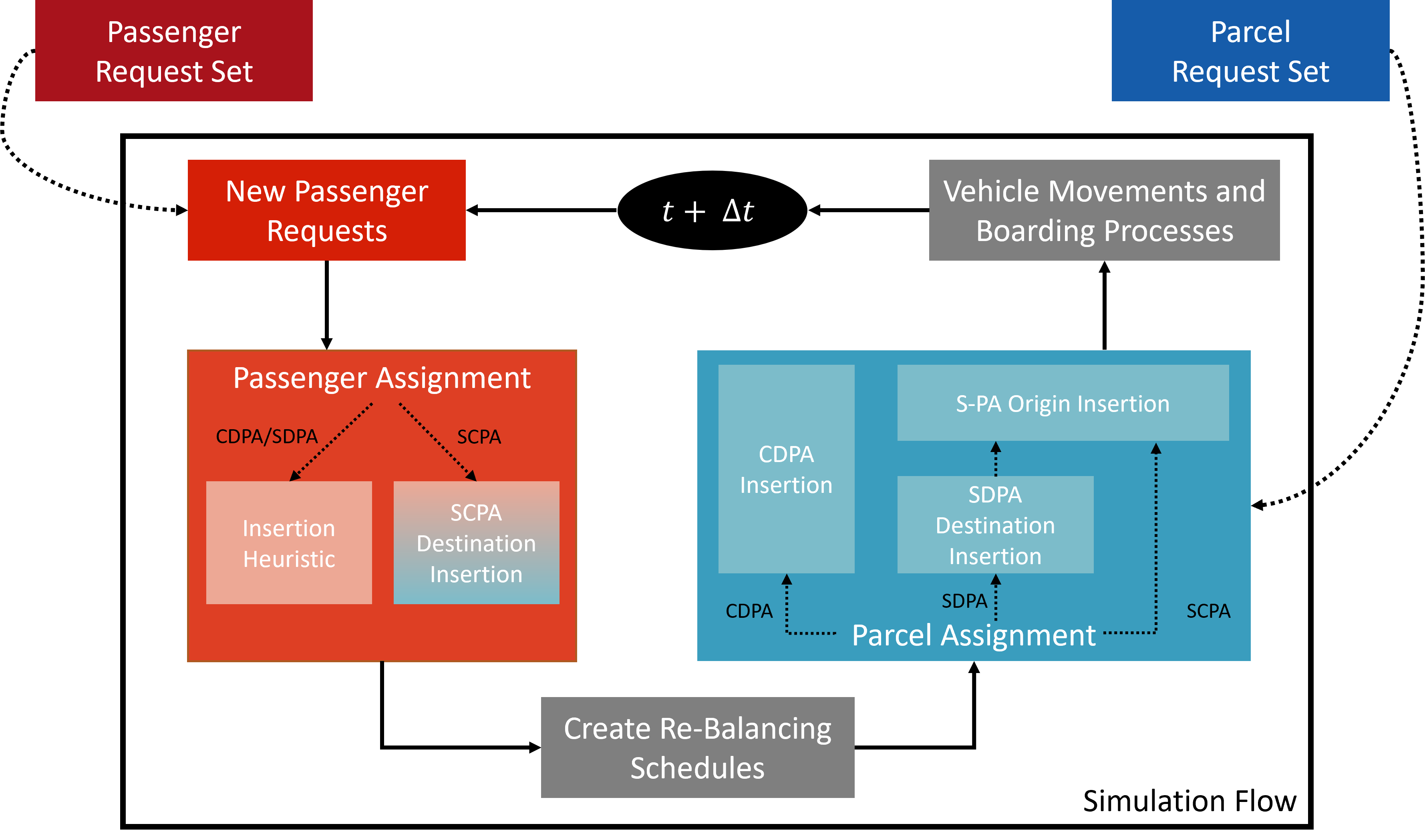}
    \caption[Caption for LOF]{Flowchart of the simulation framework with different proposed parcel assignment strategies.}
    \label{fig:highlevel_flowchart}
\end{figure}

The high-level simulation flow is shown in Figure~\ref{fig:highlevel_flowchart}. The demand for the simulation is split into a passenger request set and a parcel request set. It is assumed that the operator has access to all parcel requests for the whole simulation period and can freely decide when to serve which parcel. Passenger requests, on the other hand, are revealed to the fleet operator dynamically during the course of the simulation. Within each time step, the operator first tries to accommodate new customer requests by inserting them into current vehicle schedules. With a given frequency, namely every $T_{repo}$, re-balancing trips to distribute idle vehicles according to expected demand are computed. Based on new vehicle assignments, the decision is made to serve specific parcels. Lastly, vehicle movements and boarding processes are performed. Details on customer insertion, re-balancing, and the decision process to serve parcels are provided in the following subsections. Because the applied control strategy of ride-pooling fleets is explored in various other research, the focus of this chapter is on the description of the methodology to integrate parcel pick-up and deliveries into the existing ride-pooling control strategy. 

\subsubsection{Passenger Assignment and Re-Balancing}

In order to assign new customer requests to vehicles and corresponding schedules, a simple insertion heuristic is applied in this study. With the currently assigned schedule $\psi_k(v;R_\psi, P_\psi)$ of vehicle $v$, the pick-up and drop-off processes for a new customer request $r_i^c$ are inserted at all possible positions within the currently existing sequence of stops (drop-off must follow the pick-up stop). The new set of feasible schedules can be enumerated again and results in schedules $\psi_{\tilde{k}}(v;R_\psi \cup \{r_i^c\}, P_\psi)$ if a feasible insertion can be found. The selected vehicle $v_a$ and schedule $\psi_l$ for serving the customer request is then determined by:
\begin{equation}
    v_a, \psi_l = \argmin_{v, \psi_{\tilde{k}}} \phi(\psi_{\tilde{k}}(v;R_\psi \cup \{r_i^c\}, P_\psi)) - \phi(\psi_k(v;R_\psi, P_\psi)) ~~~ \forall v, \psi_{\tilde{k}} ~,
\end{equation}
i.e. the vehicle schedule is assigned that decreases the change in objective value the most when the new request is served. Each time a new customer requests a trip, the schedules are updated iteratively. If no solution is found by the insertion heuristic, i.e. no vehicle can serve the customer within the given time constraints, the customer leaves the system unserved.

While more sophisticated algorithms to solve the ride-pooling assignment can be found in the literature (e.g. \cite{AlonsoMora2017, Engelhardt.7292020}), using this simple insertion heuristic has the advantage that the assignment of customer and parcel requests can be decoupled in different decision processes reducing the overall complexity and thereby computational time.

To distribute idle vehicles dynamically according to expected demand in the network, a re-balancing algorithm is applied in periodical intervals of $T_{repo}$. For each zone in the network expected demand is determined and idle vehicles estimated. A parameter-free re-balancing strategy based on \cite{Pavone.2012} is applied to distribute idle vehicles by solving a minimum transport problem.

\subsubsection{Parcel Assignment}

Because no explicit time constraints for parcel pick-up and delivery are imposed in this study, different assignment strategies (in comparison to the passenger assignment) were developed to assign parcels to the vehicle schedules. While parcels could be served in times of low customer demand to increase temporal vehicle utilization, the goal of this study is to evaluate if parcel pick-up and delivery can be performed when vehicles pass by occasionally to minimize the need for additional driven fleet kilometers. Three different parcel assignment strategies are developed which will be described in the following.

\paragraph{Combined Decoupled Parcel Assignment (CDPA)} In the first assignment strategy, both the origin as well as the destination of a parcel are assigned at once. An assignment of a parcel request $r_i^p$ to vehicle $v$ is only made if the detour to add the pick-up and the drop-off into the currently assigned schedule $\psi_k(v;R_\psi, P_\psi)$ is small compared to the distance of the direct parcel route $d(o_i^p, d_i^p)$. The detour is measured by comparing the distance that has to be driven to complete the schedule, including the new parcel request $\psi_l(v;R_\psi, P_\psi \cup \{r_i^p\})$, with the distance not considering the parcel, i.e. $\psi_k(v;R_\psi, P_\psi))$. A possible assignment is identified if:
\begin{equation}
\label{eq:th_CDPA}
    d(\psi_l(v;R_\psi, P_\psi \cup \{r_i^p\})) - d((\psi_k(v;R_\psi, P_\psi))) < (1-\tau_{th}) d(o_i^p, d_i^p) ,
\end{equation}
with a threshold parameter $\tau_{th} \in \{0,1\}$ indicating the amount of detour relative to a direct route to be accepted to serve the parcel $r_i^p$. If $\tau_{th}$ approaches $1$ no detour is accepted to serve the parcel.\\
Algorithm~\ref{alg:CDPA} sketches the procedure of assigning parcels with the CDPA strategy. Let $P_u$ be the set of currently unassigned parcels and $V^{ca}$ the set of vehicles with schedules, that have been updated in the current simulation time step. The method \textit{insert}$(\psi_k(v;R_\psi, P_\psi), r_i^p)$ returns the best feasible insertion of $r_i^p$ in $\psi_k(v;R_\psi, P_\psi)$ with respect to the objective function $\phi$. For each unassigned parcel, an insertion is checked for each vehicle with an updated schedule (an insertion in other vehicles would have already been checked in previous time steps). If a new schedule fulfills the constraint of Equation \ref{eq:th_CDPA}, a candidate insertion is found. In the end, the candidate schedule with the minimum objective value is picked to be assigned. If no candidate is found, the parcel is tried to be assigned again at a later time step.

\begin{algorithm}
\caption{CDPA Insertion}
\label{alg:CDPA}
\begin{algorithmic}
\FORALL{$r_i^p \in P_u$}
    \STATE $\psi_{best} = $~None
    \STATE $v_{best} = $~None
    \FORALL {$v \in V^{ca}$}
        \STATE $\psi_{\tilde{k}}(v;R_\psi, P_\psi \cup \{r_i^p\}) = $~insert$(\psi_k(v;R_\psi, P_\psi), r_i^p)$
        \IF {$d(\psi_{\tilde{k}}(v;R_\psi, P_\psi \cup \{r_i^p\})) - d(\psi_k(v;R_\psi, P_\psi)) < (1-\tau_{th}) d(o_i^p, d_i^p)$}
            \IF{$\phi(\psi_{\tilde{k}}(v;R_\psi, P_\psi \cup \{r_i^p\}) < \phi(\psi_{best})$}
                \STATE $\psi_{best} \leftarrow \psi_{\tilde{k}}(v;R_\psi, P_\psi \cup \{r_i^p\})$
                \STATE $v_{best} \leftarrow v$
            \ENDIF
        \ENDIF
    \ENDFOR
    \IF {$v_{best} \neq None$}
        \STATE assignSchedule$(v_{best}, \psi_{best})$
        \STATE $P_u \leftarrow  P_u \setminus \{r_i^p\}$
    \ENDIF
\ENDFOR
\end{algorithmic}
\end{algorithm}

\begin{algorithm}[!t]
\caption{S-PA Origin Insertion}
\label{alg:SPA_O}
\begin{algorithmic}
\FORALL{$r_i^p \in P_u$}
    \STATE $\psi_{best} = $~None
    \STATE $v_{best} = $~None
    \FORALL {$v \in V^{ca}$}
        \STATE $\psi_{\tilde{k}}(v;R_\psi, P_\psi \cup \{o_i^p\}) = $~insertOrigin$(\psi_k(v;R_\psi, P_\psi), r_i^p)$
        \IF {$d(\psi_{\tilde{k}}(v;R_\psi, P_\psi \cup \{o_i^p\})) - d(\psi_k(v;R_\psi, P_\psi)) < (1-\tau_{th}) d(o_i^p, d_i^p)/2$}
            \IF{$\phi(\psi_{\tilde{k}}(v;R_\psi, P_\psi \cup \{o_i^p\}) < \phi(\psi_{best})$}
                \STATE $\psi_{best} \leftarrow \psi_{\tilde{k}}(v;R_\psi, P_\psi \cup \{r_i^p\})$
                \STATE $v_{best} \leftarrow v$
            \ENDIF
        \ENDIF
    \ENDFOR
    \IF {$v_{best} \neq $~None}
        \STATE assignSchedule$(v_{best}, \psi_{best})$
        \STATE $P_u \leftarrow  P_u \setminus \{r_i^p\}$
        \STATE $P_{v_{best}}^a \leftarrow P_{v_{best}}^a \cup \{r_i^p\}$
    \ENDIF
\ENDFOR
\end{algorithmic}
\end{algorithm}

\begin{algorithm}[!b]
\caption{SDPA Destination Insertion}
\label{alg:SDPA}
\begin{algorithmic}
\FORALL {$v \in V^{ca}$}
    \STATE $\psi_{best} = $~None
    \STATE $r_{best} = $~None
    \FORALL{$r_i^p \in P_v^a$}
        \STATE $\psi_{\tilde{k}}(v;R_\psi, P_\psi \cup \{r_i^p\}) = $~insertDestination$(\psi_k(v;R_\psi, P_\psi \cup \{o_i^p\}), r_i^p)$
        \IF {$d(\psi_{\tilde{k}}(v;R_\psi, P_\psi \cup \{r_i^p\})) - d(\psi_k(v;R_\psi, P_\psi \cup \{o_i^p\})) < (1-\tau_{th}) d(o_i^p, d_i^p)/2$}
            \IF{$\phi(\psi_{\tilde{k}}(v;R_\psi, P_\psi \cup \{r_i^p\})) < \phi(\psi_{best})$}
                \STATE $\psi_{best} \leftarrow \psi_{\tilde{k}}(v;R_\psi, P_\psi \cup \{r_i^p\})$
                \STATE $r_{best} \leftarrow r_i^p$
            \ENDIF
        \ENDIF
    \ENDFOR
    \IF {$r_{best} \neq $~None}
        \STATE assignSchedule$(v, \psi_{best})$
        \STATE $P_v^a \leftarrow  P_v^a \setminus r_i^p$
    \ENDIF
\ENDFOR
\end{algorithmic}
\end{algorithm}

\paragraph{Subsequent Decoupled Parcel Assignment (SDPA)} The idea of the second assignment strategy is that -- because no time constraints are imposed on the parcel drop-off -- the decision on when to drop-off the parcel does not have to be made when the decision of the parcel pick-up is made. This assumption can be made under the condition that the loaded parcels do not significantly affect the energy consumption of the vehicles. Therefore, the decision to pick-up a parcel is separated from the decision to drop-off a parcel.
The decision to pick-up a parcel is taken similar to the CDPA strategy and summarized in Algorithm~\ref{alg:SPA_O}. The differences to Algorithm~\ref{alg:CDPA} can be summarized as follows: firstly, only the insertion of the origin $o_i^p$ is tested for parcel request $r_i^p$ by the method \textit{insertOrigin}, secondly, the possible assignment is identified if:
\begin{equation}
    d(\psi_l(v;R_\psi, P_\psi \cup \{o_i^p\})) - d((\psi_k(v;R_\psi, P_\psi))) < (1-\tau_{th}) d(o_i^p, d_i^p)/2 .
\end{equation}
Note that the threshold is divided by $2$ (compared to CDPA) to account for an even split of the overall detour for parcel pick-up and drop-off. If a feasible insertion of the origin of $r_i^p$ is assigned to vehicle $v$, $r_i^p$ is added to the set $P_v^a$ to keep track of assigned parcel pick-ups for each vehicle.\\
A similar approach is chosen to assign the parcel drop-off in any later simulation time step and is sketched in Algorithm \ref{alg:SDPA}. For all vehicles with scheduled pick-ups or on-board parcels, the insertions of drop-offs for all parcels $r_i^p \in P_v^a$ is checked. A possible assignment is considered if:
\begin{equation}
    d(\psi_l(v;R_\psi, P_\psi \cup \{r_i^p\})) - d((\psi_k(v;R_\psi, P_\psi \cup \{o_i^p\}))) < (1-\tau_{th}) d(o_i^p, d_i^p)/2 .
\end{equation}
Thereby, $\psi_l(v;R_\psi, P_\psi \cup \{r_i^p\})$ refers to the schedule including origin $o_i^p$ and destination $d_i^p$ of parcel $r_i^p$.\\
In the simulation, at first possible insertions of parcel drop-offs are tested for vehicles with an updated schedule, then possible pick-up insertions are created for parcel pick-ups analogously to the comprehensive parcel assignment strategy. This strategy has the downside that it is not guaranteed to find a drop-off for each parcel until the end of the simulation. However, parcels should not remain on board the vehicles until the end. Hence, when a certain simulation time $T_p^{max}$ is exceeded, all remaining on-board parcels are scheduled to be dropped off by iteratively inserting them into the current vehicle schedule.

\begin{algorithm}[!b]
\caption{SCPA Destination Insertion}
\label{alg:SCPA}
\begin{algorithmic}
\FORALL{$r_i^c \in R_t^{new}$}
    \STATE $\psi_{best,u} = $~None
    \STATE $v_{best,u} = $~None
    \FORALL{$v \in V$}
        \STATE $\psi_{\tilde{k}}(v;R_\psi \cup \{r_i^c\}, P_\psi) = $~insert$(\psi_k(v;R_\psi, P_\psi), r_i^c)$
        \IF{$\phi(\psi_{\tilde{k}}(v;R_\psi \cup \{r_i^c\}, P_\psi)) < \phi(v_{best,u})$}
            \STATE $\psi_{best,u} \leftarrow \psi_{\tilde{k}}(v;R_\psi \cup \{r_i^c\}, P_\psi)$
            \STATE $v_{best,u} \leftarrow v$
        \ENDIF
    \ENDFOR
    \STATE $\psi_{best} = \psi_{best,u}$
    \STATE $v_{best} = v_{best,u}$
    \STATE $r_{best} = $~None
    \FORALL{$v \in V$}
        \STATE $\psi_{\tilde{k}}(v;R_\psi \cup \{r_i^c\}, P_\psi) = $~insert$(\psi_k(v;R_\psi, P_\psi), r_i^c)$
        \FORALL{$r_i^p \in P_v^a$}
            \STATE $\psi_{l}(v;R_\psi \cup \{r_i^c\}, P_\psi \cup \{r_i^p\}) = $~insertDestination$(\psi_{\tilde{k}}(v;R_\psi \cup \{r_i^c\}, P_\psi), r_i^p)$ 
            \IF {$d(\psi_{l}(v;R_\psi \cup \{r_i^c\}, P_\psi \cup \{r_i^p\})) - d(\psi_{best,u}) < (1-\tau_{th}) d(o_i^p, d_i^p)/2$}
                \IF{$\phi(\psi_{l}(v;R_\psi \cup \{r_i^c\}, P_\psi \cup \{r_i^p\})) < \phi(\psi_{best})$}
                    \STATE $\psi_{best} \leftarrow \psi_l(v;R_\psi, P_\psi \cup \{r_i^p\})$
                    \STATE $r_{best} \leftarrow r_i^p$
                    \STATE $v_{best} \leftarrow v$
                \ENDIF
            \ENDIF
        \ENDFOR
    \ENDFOR
    \IF {$v_{best} \neq $~None}
        \STATE assignSchedule$(v, \psi_{best})$
        \IF{$r_{best} \neq $~None}
            \STATE $P_{v_{best}}^a \leftarrow  P_{v_{best}}^a \setminus \{r_{best}\}$
        \ENDIF
    \ENDIF
\ENDFOR
\end{algorithmic}
\end{algorithm}

\paragraph{Subsequent Coupled Parcel Assignment (SCPA)} In the last assignment strategy, the decision of assigning the drop-off of a parcel is again made independently from the decision of assigning the pick-up. While the decision of assigning the pick-up remains the same compared to the subsequent independent parcel assignment strategy (Algorithm \ref{alg:SPA_O}), the decision of assigning the drop-off is coupled to the passenger assignment. The idea is to assign passengers and thereby new vehicle schedules passing by the drop-off location of on-board parcels. By that, the possible solution space for parcel drop-off assignments increases. To this end, the formulation of the passenger assignment is revisited.\\
Let $R_t^{new}$ be the set of new customer requests in time step $t$. In the first step, the best possible solution for just serving a new customer request $r_i^c \in R_t^{new}$ is calculated using the same insertion heuristic as for the passenger assignment. The resulting schedule $\psi_l(v_a) = \psi_l(v;R_\psi \cup \{r_i^c\}, P_\psi)$ is used as a benchmark for the decision to assign a parcel drop-off. In a second step, instead of inserting a parcel drop-off into the overall best solution $\psi_l(v_a)$, an insertion of each on-board parcel requests $r_i^p$ is tried for feasible schedules in combination with the new request $r_i^c$, resulting in the schedules $\psi_{\tilde{l}}(v;R_\psi \cup \{r_i^c\}, P_\psi \cup \{r_i^p\}) \forall v \in V_{r_i^c}$. A possible assignment of the parcel is found if:
\begin{equation}
    d(\psi_{\tilde{l}}(v;R_\psi \cup \{r_i^c\}, P_\psi \cup \{r_i^p\})) - d(\psi_l(v_a;R_\psi \cup \{r_i^c\}, P_\psi)) < (1-\tau_{th}) d(o_i^p, d_i^p)/2 ,
\end{equation}
i.e. the driven distance of the best possible solution without parcel delivery is only increased by at most a threshold factor compared to the direct distance of the inserted parcels. If multiple of these options exist, the vehicle schedule minimizing the objective $\phi$ is assigned. If none of these options exist, only the best schedule for serving the customer is assigned. The corresponding logic is sketched in Algorithm \ref{alg:SCPA}. For reasons of clarity and comprehensibility, not the computationally most efficient version of the algorithm is depicted in Algorithm \ref{alg:SCPA}. For example, all solutions from the first customer insertion can be stored in a list, which makes the re-computation of the insertion in the second loop over vehicles redundant.

\subsection{Case Study for Munich, Germany}

\begin{figure}[!b]
    \centering
    \includegraphics[width=0.7\textwidth]{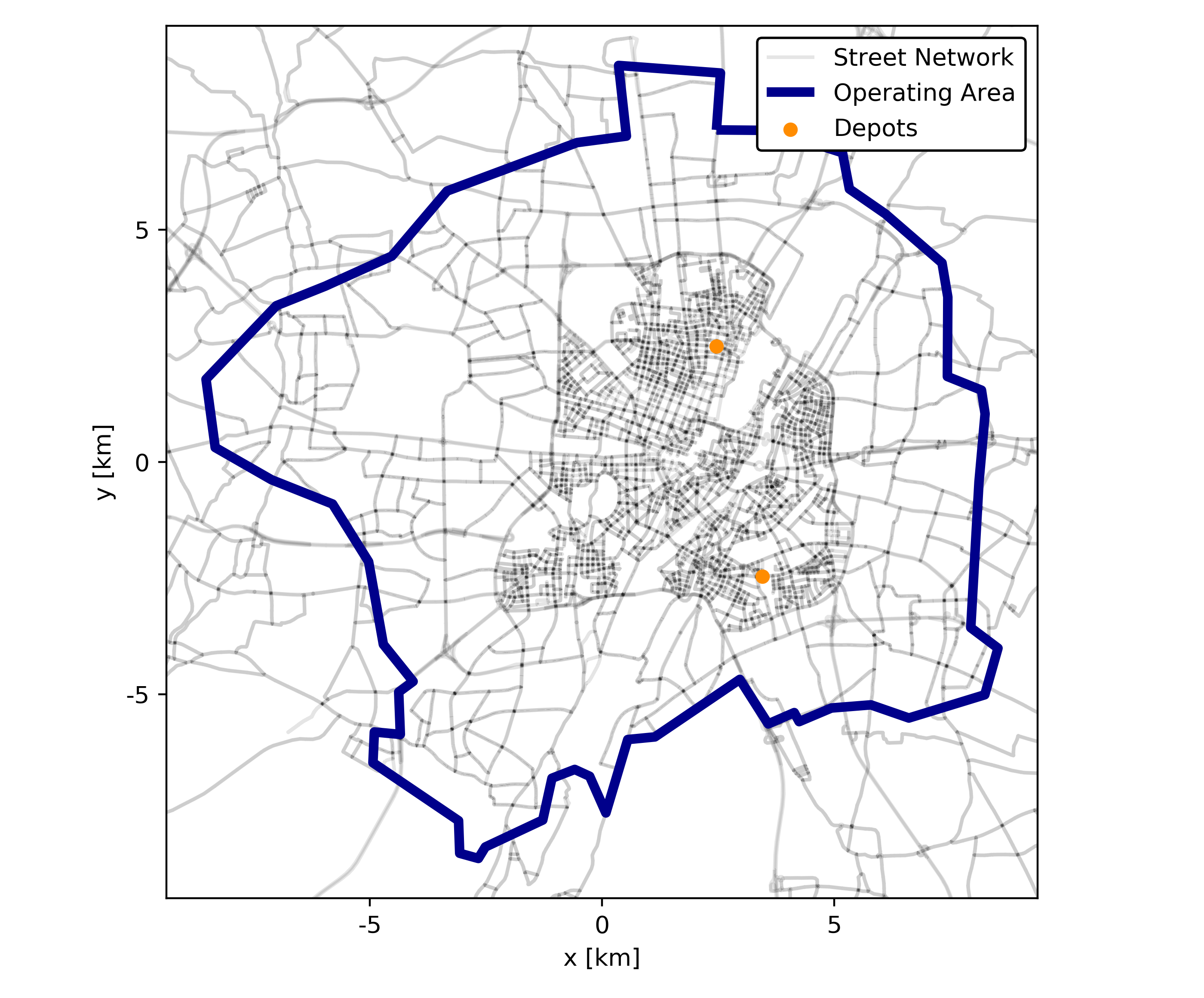}
    \caption{Street network, operating area and depots for the modelled RPP service in the case study of Munich, Germany.}
    \label{fig:ODM-Service-Area}
\end{figure}

The proposed framework is applied and evaluated for a case study in Munich, Germany. The considered service area of the MoD operator extends almost to the freeway ring, which surrounds the city center. The street network $G=(N,E)$ with edge travel times for each hour of a usual working day is extracted from a calibrated microscopic traffic simulation described in \cite{F.Dandl.2017}. The street network and operating area are shown in Figure~\ref{fig:ODM-Service-Area}.

Passenger demand for the MoD service is created by sampling from private vehicle trip OD-matrices, extracted from the same microscopic traffic simulation. 24 matrices for each hour of the day are available, which contain around one million trips starting and ending within the service area. Poisson processes are used to sample requests for the MoD service with Poisson rates defined by the corresponding OD-entries times a penetration factor. In this study, a penetration of 5\% is applied, which can be interpreted as a MoD service replacing 5\% of the private vehicle trips within the operating area resulting in approximately 50,000 trips per day. Origin and destination nodes for the sampled requests are matched randomly onto intersection nodes within associated zones defined by the OD-matrices. Using different random seeds for the sampling process, three different sets of requests are created and used for the simulations.

One month of real-world parcel shipment data of a local logistics provider was available to create parcel demand for the RPP service. The data includes the date of delivery, the local depot the parcel has been delivered from, and a destination address. In the first step, the destination address is converted into coordinates using the open-source geo-coding API 'Nominatim' relying on publicly available OpenStreetMap data. All deliveries outside of the operating area of the MoD service are removed resulting in 56,000 parcels within the operating area. Using the coordinates, parcel delivery destinations are matched onto the nearest intersection node in the network. Maximally $c_v^p$ parcels with same origin-destination relation are aggregated into the same parcel request. Because no information about the size of the parcels is given in the data, the size of a parcel request is set by the number of aggregated parcels. All parcels are shipped from two depots, both positioned outside of the MoD operating area (in the north and east of Munich). It is assumed that if a RPP service as represented in this study is introduced, corresponding depots have to be implemented within the operating area. Therefore, two new depots in the northern and eastern part of the city are defined and shown in Figure~\ref{fig:ODM-Service-Area}. Parcels that are shipped from the original northern (eastern) depot are assumed that they have to be delivered from the introduced inner-city northern (eastern) depot. One goal of the study is to observe the system boundaries, i.e. which amount of parcel demand can be served with given passenger demand. Therefore, the parcel demand data for the whole month is used as input for the simulations. To be able to vary the ratio between passenger and parcel demand is sub-sampled to shares of 1\% to 50\% of the overall parcel demand. In the base scenario, a 10\% sub-sample of the parcel demand is used resulting in a ratio of approximately 1 parcel to 10 passenger requests to represent a service with a priority on serving passenger demand. Analogously to the passenger demand, three different sets of parcel requests are created using different random seeds within the sub-sampling process.

Contrary to the simulations modeling the RPP service, the \textit{Status Quo} is modeled with two independent vehicle fleets serving as a baseline to evaluate the efficiency of integrating parcel delivery into the MoD service. The first vehicle fleet corresponds to the MoD service without delivering any parcels. Hereby, the described simulations are conducted without any parcel demand. The second fleet corresponds to the pure logistic service. Hereby, vehicles are placed initially at each of the two depots. Parcels for the corresponding demand scenario are inserted iteratively into their schedules, minimizing the driven distance including their return to the depot at the end of the route. The aggregate driven distance within these schedules is used to approximate the fleet vehicle kilometer by the logistic service.

Within the simulation, the service parameters describing customer maximum waiting time $t_{max}^{wait}$ is set to $10$~min while the maximum detour factor $\Delta$ is set to $40$\%. The fleet is operated with a capacity of $c_v^c = 4$ passengers and $c_v^p = 8$ parcels. In the \textit{Status Quo} scenario, it is assumed that the logistics provider operates trucks with a capacity of 100 parcels \cite{DHL2019}. Re-balancing trips are performed every $T_{repo} = 15$~min and based on forecasts according to the zones and values of the corresponding entries in the OD-matrices times the penetration factor of 5\%. The fleet size of the RPP operator for the following simulations is determined by first performing multiple simulations without parcel demand and varying fleet size. Finally, a fleet size of 600 vehicles is chosen that allows a service rate of approximately 90\% served customers.

First simulations are performed for all parcel assignment strategies (CDPA, SCPA, SDPA) as well as for \textit{Full RPP Integration} and \textit{Moderate RPP Integration}. Within these simulations, the influence of the assignment threshold parameter $\tau_{th}$ on various service Key-Performance-Indicators (KPIs) is evaluated. In a second step, simulations with varying parcel demand penetration ranging from $0$\% to $50$\% of the overall parcel demand data set are performed while keeping $\tau_{th}$ constant to evaluate the number of parcels that can be accommodated by the RPP service keeping the passenger demand fixed. Within the SDPA and SCPA strategy remaining on board parcels are actively delivered after $T_p^{max} = 10$pm.

\section{Results}
Results of the simulation from the presented case study for Munich, Germany are presented in the following. In the upcoming two subsections influences on customer KPIs and operator KPIs are evaluated based on the different proposed RPP service integration and parcel assignment strategies with a varying threshold parameter $\tau_{th}$. In the third subsection, effects resulting from different parcel demand levels are evaluated and compared to the \textit{Status Quo}.

\subsection{Influence on Customer KPIs}

A crucial question for the success of the proposed RPP service will be, whether the ride-pooling fleet can still ensure sufficient service quality for passengers despite the additional transport of parcels. The attitude in this paper is that logistics services are subordinated to passenger requests. In this context, for the scenario \textit{Moderate RPP Integration} the pick-up or delivery of parcels is only allowed while no passengers are on board, while this constraint is lifted for the scenario \textit{Full RPP Integration}. Nevertheless, time-constraints regarding passenger pick-up and maximum detour, described in the beginning of this chapter, have to be fulfilled in either scenario ensuring a certain quality of service.

\begin{figure}
     \centering
     \begin{subfigure}[b]{0.49\textwidth}
         \caption{Waiting Time}
         \includegraphics[width=\textwidth]{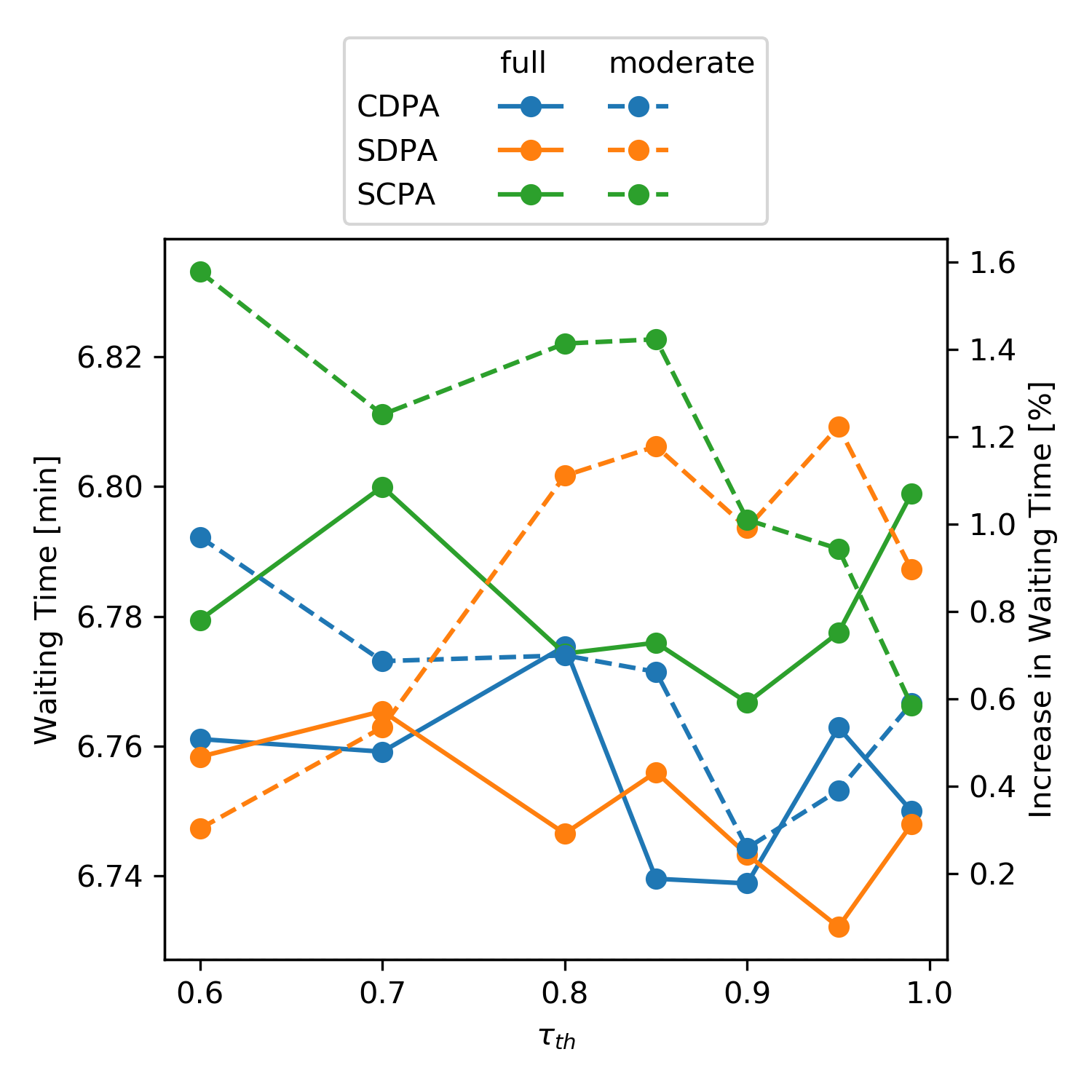}
         \label{fig:calib_wt}
     \end{subfigure}
     \hfill
     \begin{subfigure}[b]{0.49\textwidth}
         \caption{Travel Time}
         \includegraphics[width=\textwidth]{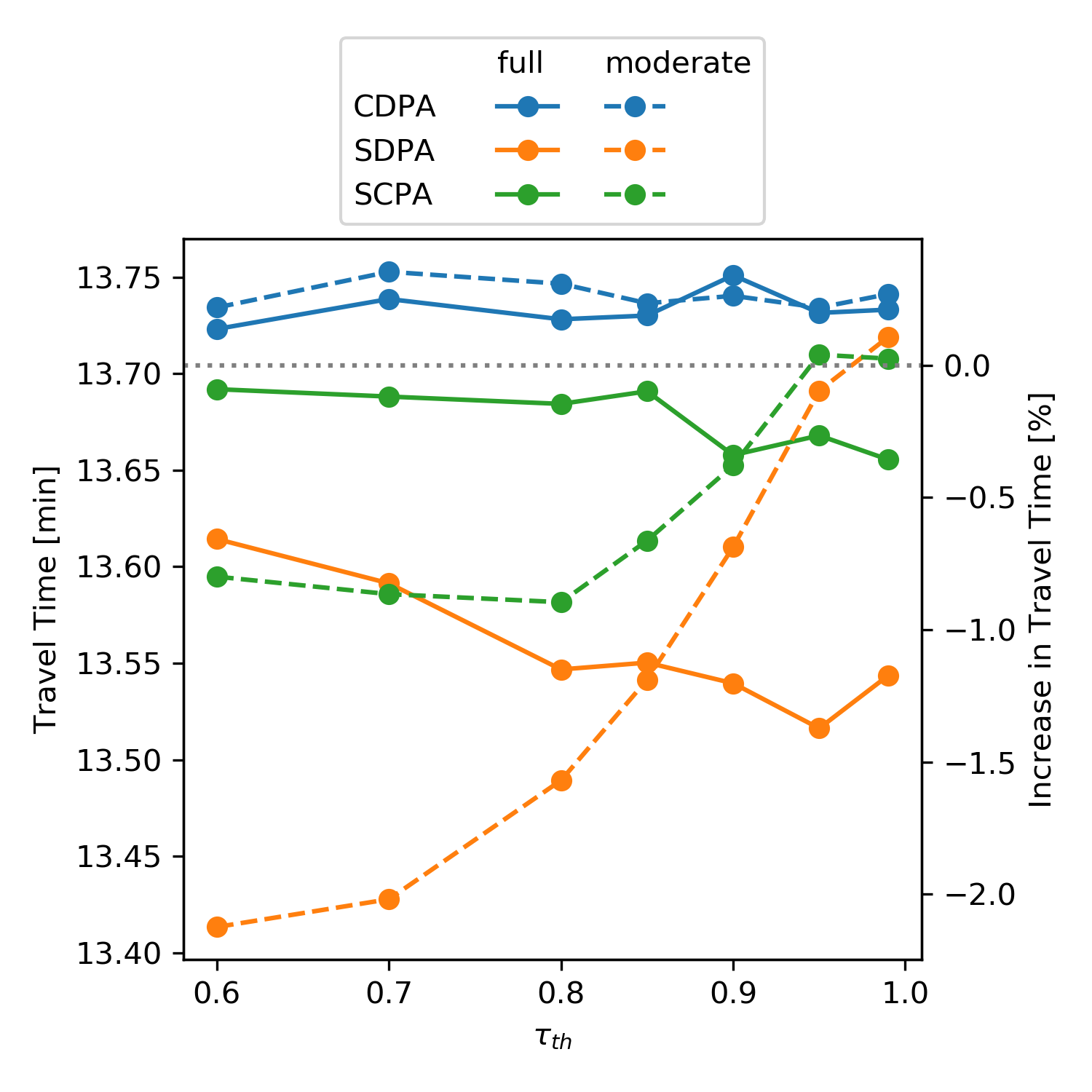}
         \label{fig:calib_tt}
     \end{subfigure}
        \caption{Impact of threshold parameter on waiting time and travel time.}
        \label{fig:wt_tt}
\end{figure}

The quality of the mobility service from the customer's perspective is evaluated by average customer waiting and travel times. The simulation results show, that waiting and travel times of the customers are rarely influenced by the integration of freight transport into the MoD service. Figure~\ref{fig:calib_wt} shows, that the customer's waiting times for all simulation scenarios and the different assignment thresholds ($\tau_{th}$) are maximally increased by 2\%. Looking closer at the blue lines, which indicate the CDPA strategy for the full (solid line) and moderate (dashed line) scenario, one can observe that increasing values for $\tau_{th}$ tend to result in lower waiting times, as the allowed detour for pick-up and drop-off of parcels becomes smaller. In the cases of the SCPA and SDPA strategies, the trend seems to be less compliant. The influences on customer's travel times are shown in Figure~\ref{fig:calib_tt}. It becomes apparent, that similar to the waiting times, the travel times are also only marginally influenced by the additional transport of freight in the MoD system. Looking at the blue lines for the CDPA strategy the increase in travel time is constantly between 0.1\% and 0.5\% for the investigated assignment thresholds ($\tau_{th}$). Taking into account the SCPA and SDPA strategies it becomes clear that especially for small assignment thresholds a considerable decrease in travel times can be observed. 

\begin{figure}
     \centering
     \begin{subfigure}[b]{0.89\textwidth}
         \caption{Absolute Waiting Time}
         \includegraphics[width=\textwidth]{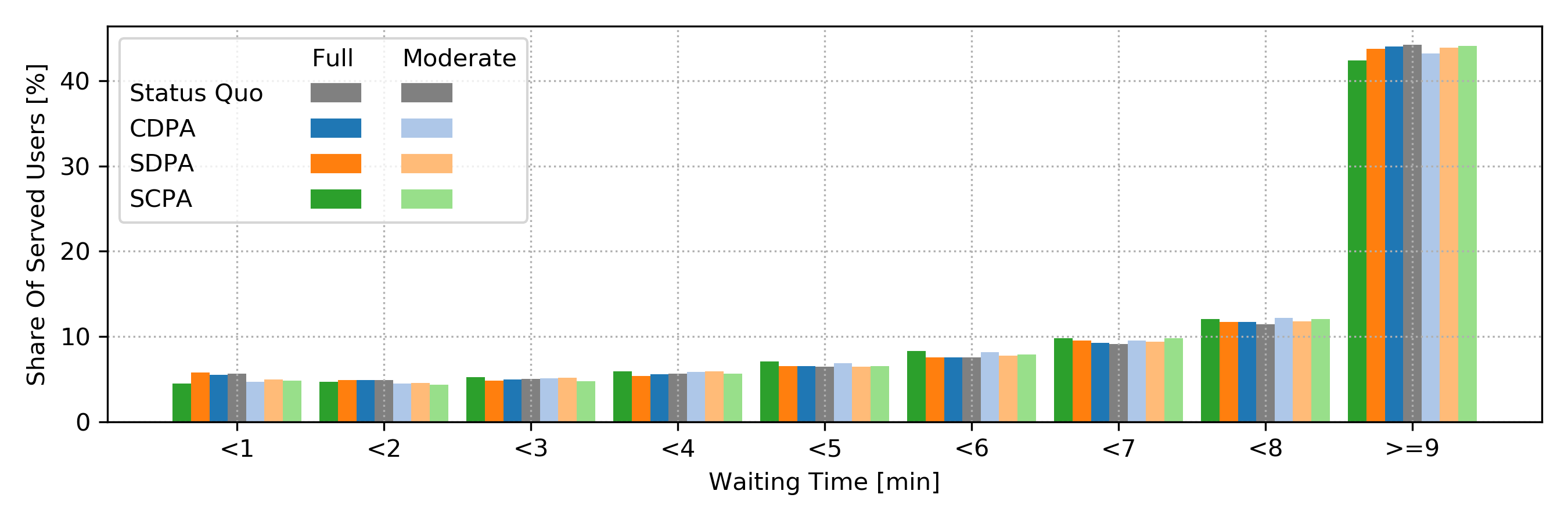}
         \label{fig:abs_wt_dist}
     \end{subfigure}
     \hfill
     \begin{subfigure}[b]{0.89\textwidth}
         \caption{Absolute Travel Time}
         \includegraphics[width=\textwidth]{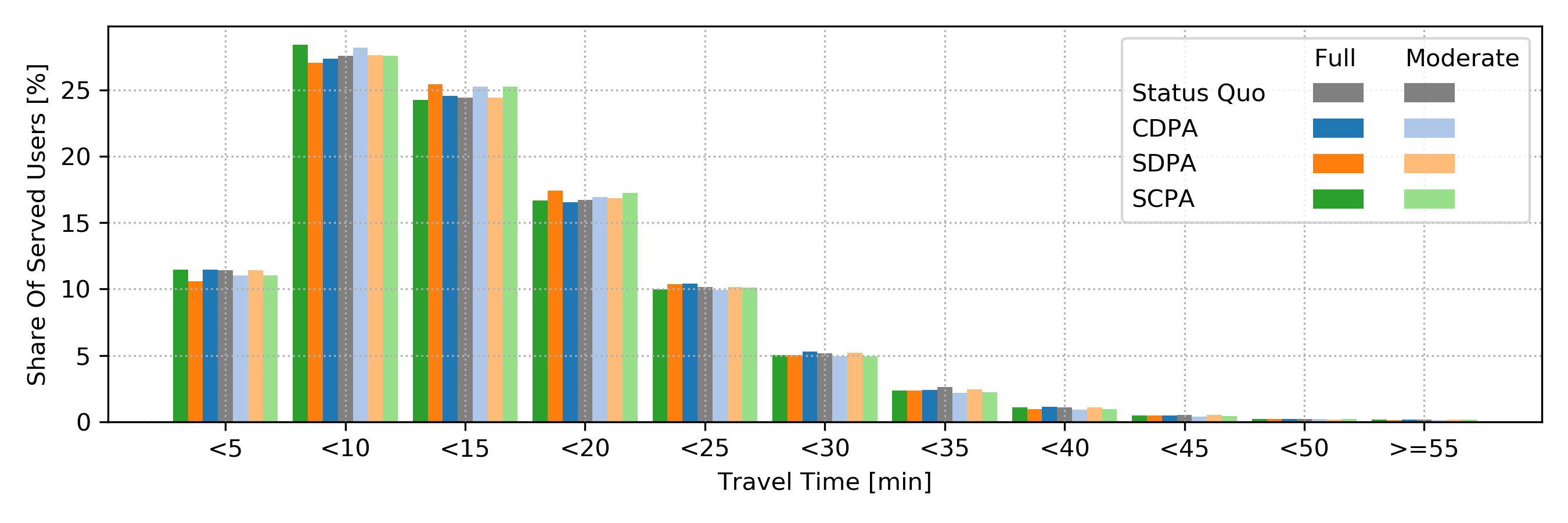}
         \label{fig:abs_tt_dist}
     \end{subfigure}
     \hfill
     \begin{subfigure}[b]{0.89\textwidth}
         \caption{Waiting Time - Change to \textit{Status Quo}}
         \includegraphics[width=\textwidth]{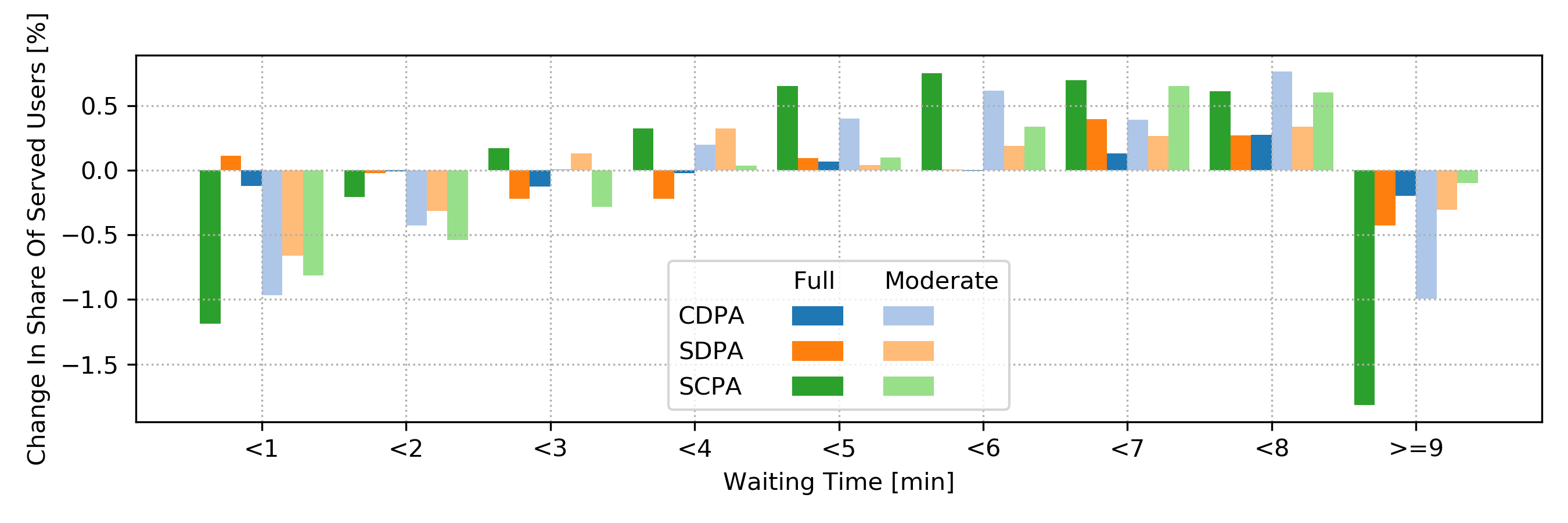}
         \label{fig:res_wt_dist}
     \end{subfigure}
    \hfill
     \begin{subfigure}[b]{0.89\textwidth}
         \caption{Travel Time - Change to \textit{Status Quo}}
         \includegraphics[width=\textwidth]{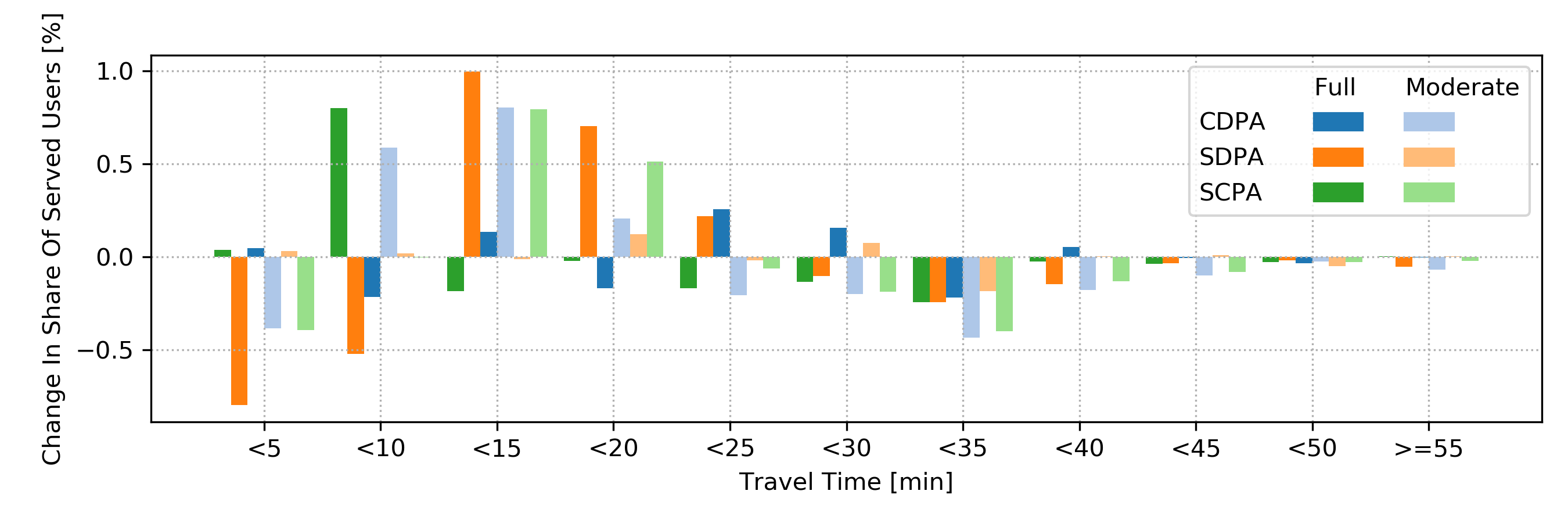}
         \label{fig:res_tt_dist}
     \end{subfigure}
        \caption{Waiting and travel time distribution of served customers within different RPP assignment strategies. $\tau_{th} = 0.80$ is considered in all scenarios shown. (fleet size = 600)}
        \label{fig:tt_wt_dist}
\end{figure}

Figure~\ref{fig:abs_wt_dist} and Figure~\ref{fig:abs_tt_dist} show the absolute waiting and travel time distribution of the customers for the different RPP scenarios and the assignment strategies. Generally, the customer assignment tends to insert new passengers with waiting times close to the maximum allowed waiting time of $10$~min. It once again becomes clear, that the difference in waiting and travel times is relatively small comparing the \textit{Status Quo} to the RPP scenarios. This means that RPP has only little negative impact on the service quality of the represented MoD provider. Looking at Figure~\ref{fig:res_wt_dist} and Figure~\ref{fig:res_tt_dist} the relative changes compared to the base scenario (\textit{Status Quo}) of the MoD ride-pooling service in terms of waiting and travel time become easier to identify. Especially, the ratio of passengers experiencing low waiting times (up to 3 minutes), resulting from preceding parcel pick-ups or drop-offs, tends to decrease. This is especially the case for the moderate integration because parcel insertion can mainly be done within passenger approaches of the vehicles, while no customer is on board yet. The decrease of passengers experiencing waiting times close to the maximum of $10$~min might be an effect of longer assigned schedules by the additional parcel service, decreasing the probability of finding a feasible insertion. For the travel time distribution, a similar trend can be observed. Short and long travel times tend to be slightly increased, whereas mid-range travel times (around 15 min) rather experience a small decrease. This indicates the trend that parcels are mainly inserted in short to medium-long trips. 

Looking closer at the difference between the \textit{Full RPP Integration} and \textit{Moderate RPP Integration} scenarios one can observe that from the customers' perspective, i.e. waiting and travel times, no big advantage is gained by limiting the parcel pick-up and drop-off to times, when no customer is on board. The detours for parcel pick-ups and drop-offs, which have to be accepted by the passengers, seem to get compensated by the additional freedom within the creation of the vehicle schedules.

\subsection{Influence on Operator KPIs}

\begin{figure}[!ht]
     \centering
     \begin{subfigure}[b]{0.49\textwidth}
         \caption{Served Parcels}
         \includegraphics[width=\textwidth]{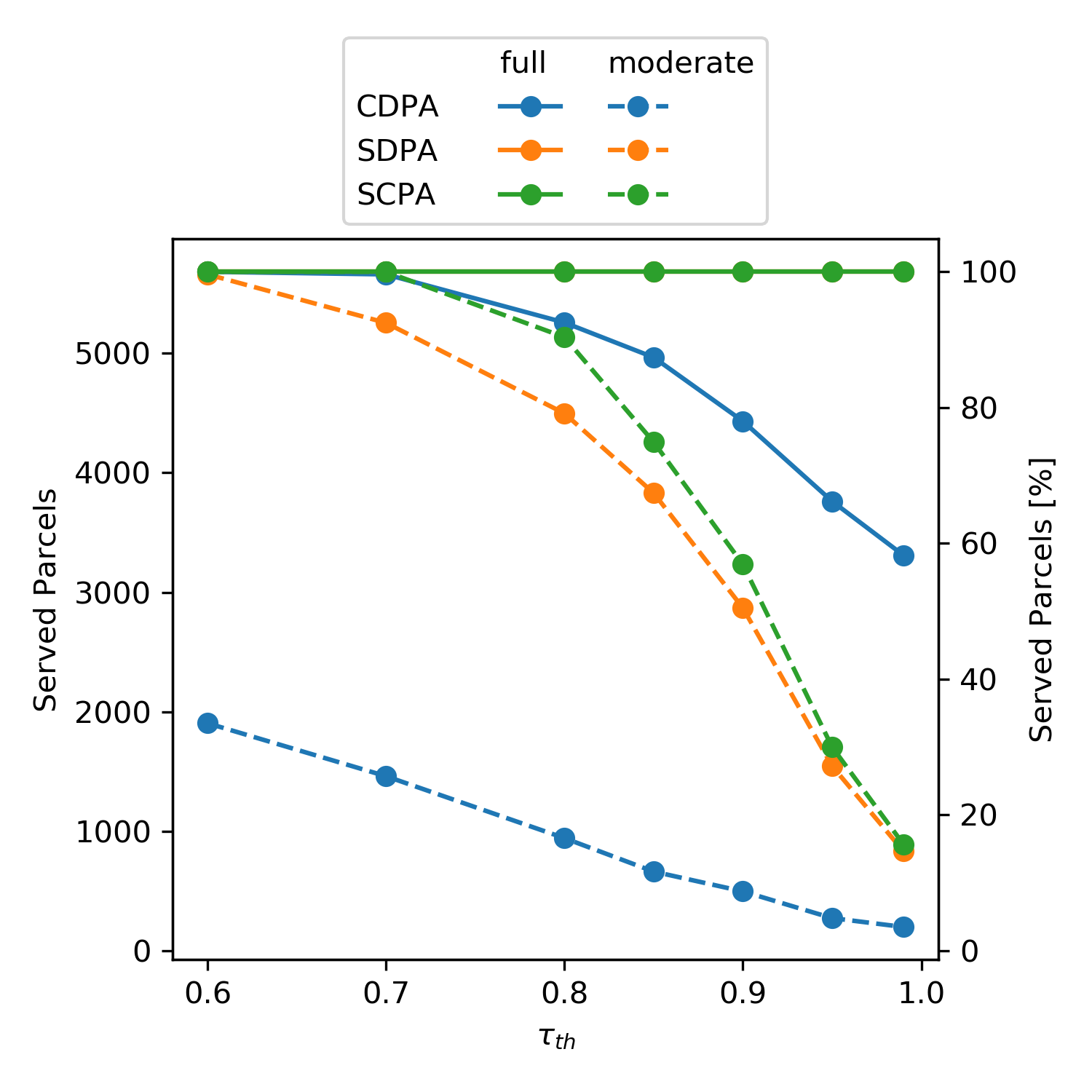}
         \label{fig:calib_served_parcels}
     \end{subfigure}
     \hfill
     \begin{subfigure}[b]{0.49\textwidth}
         \caption{Served Customers}
         \includegraphics[width=\textwidth]{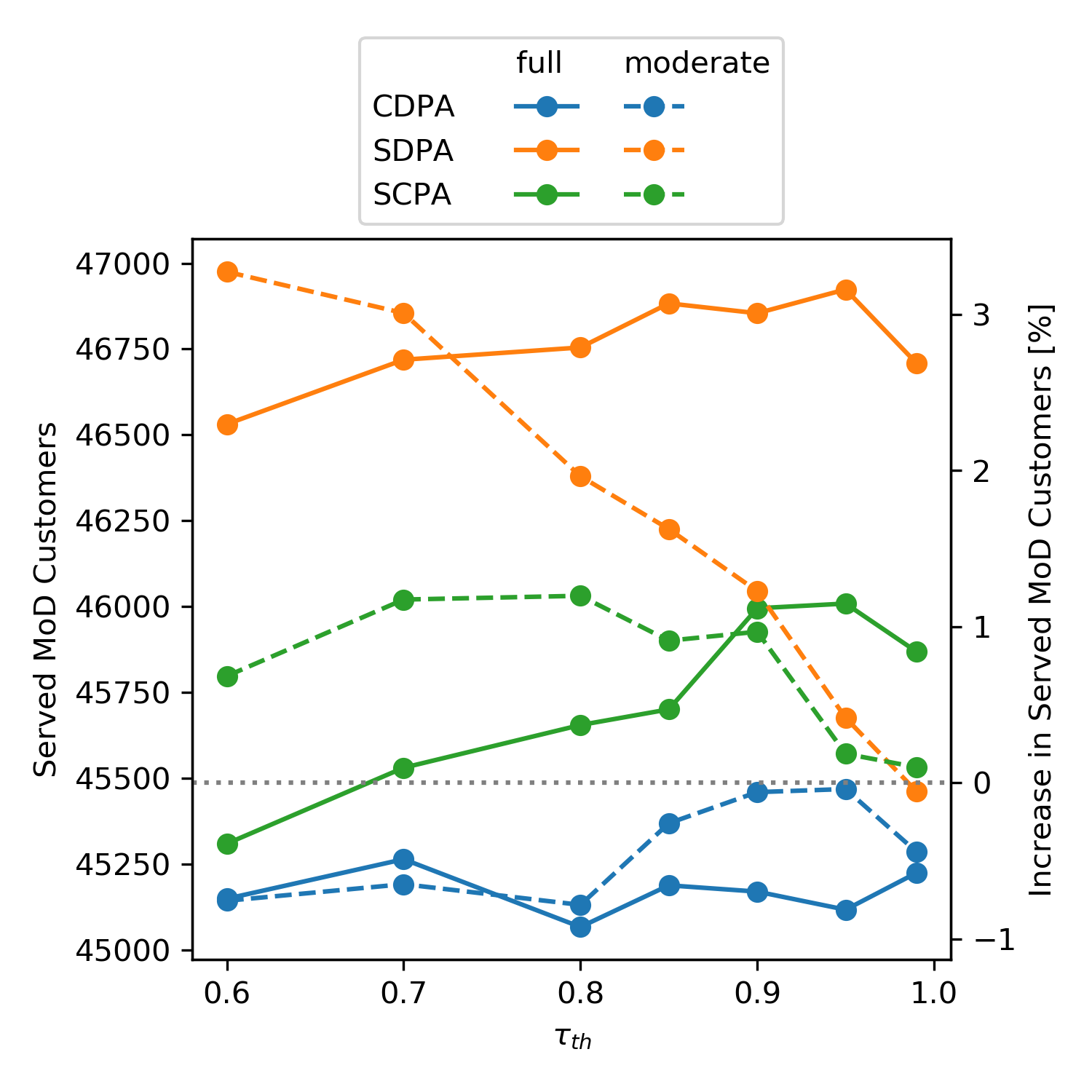}
         \label{fig:calib_served_persons}
     \end{subfigure}
     \hfill
     \begin{subfigure}[b]{0.49\textwidth}
         \caption{Fleet KM}
         \includegraphics[width=\textwidth]{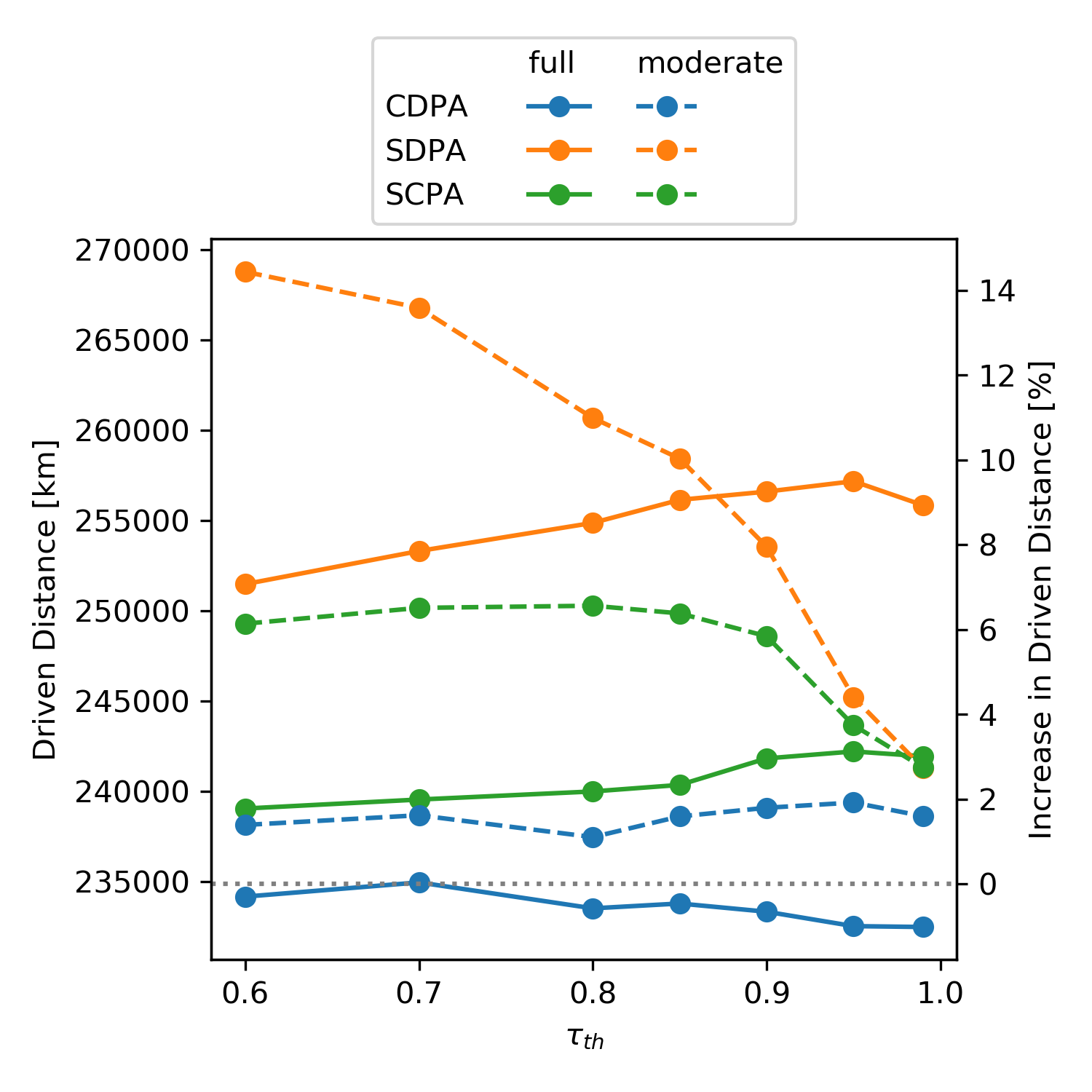}
         \label{fig:calib_vkm}
     \end{subfigure}
     \hfill
     \begin{subfigure}[b]{0.49\textwidth}
         \caption{Fleet Utilization}
         \includegraphics[width=\textwidth]{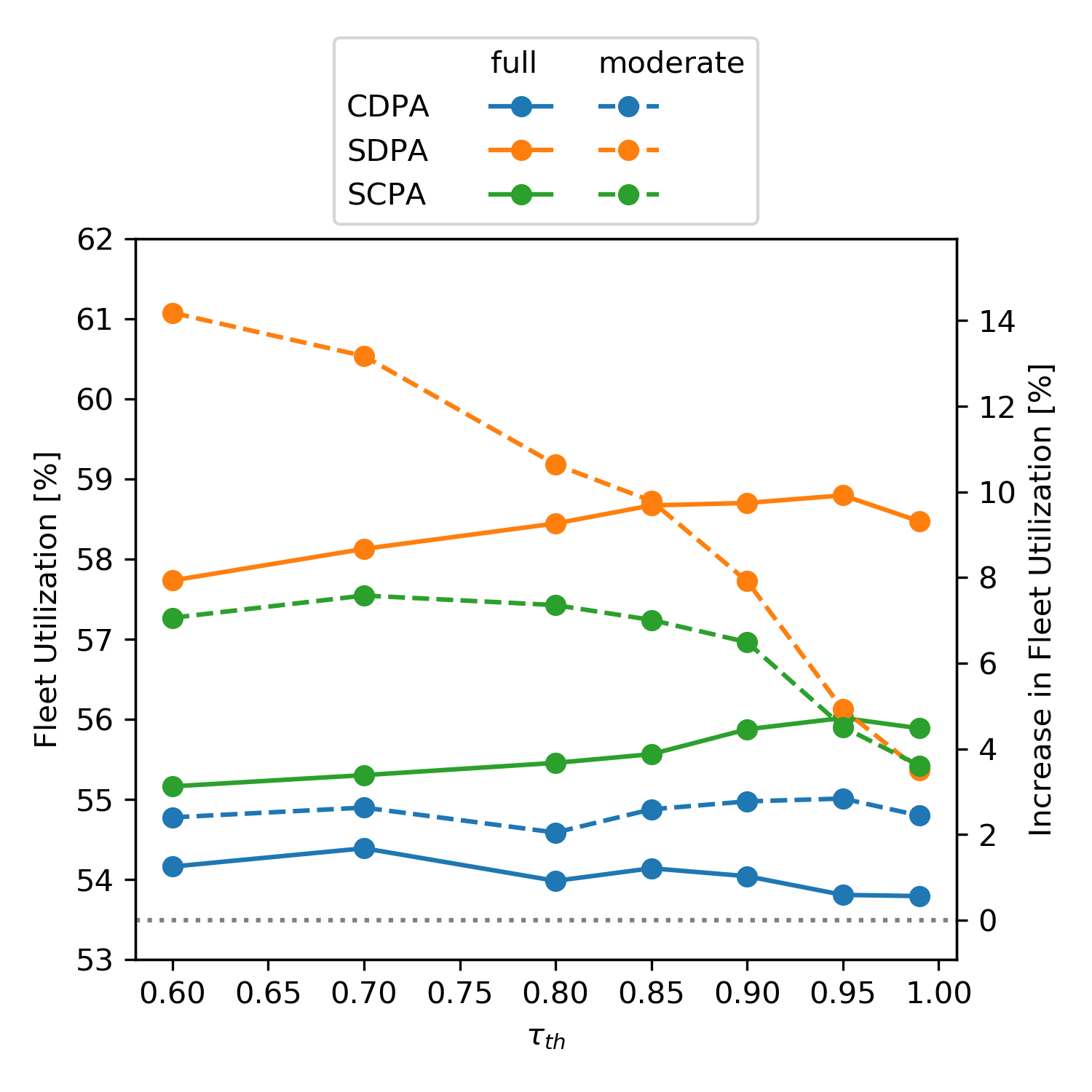}
         \label{fig:calib_util}
     \end{subfigure}
        \caption{Impact of threshold parameter on the number of served parcels, served persons, fleet kilometers traveled and fleet utilization.}
        \label{fig:calib_th}
\end{figure}

In addition to the attractiveness of a mobility service for the customer, the operator's perspective plays a decisive role for success. It is important for the provider of a MoD ride-pooling service, that the vehicle fleet can be operated efficiently even after the logistics service has been integrated.
Figure~\ref{fig:calib_served_parcels} shows the number of served parcels given assignment thresholds ($\tau_{th}$) between 0.6 and 1.0. One can observe that generally, the number of served parcels declines with an increasing assignment threshold, resulting from more stringent pick-up or drop-off constraints for a parcel in a given schedule. However, the CDPA reacts more sensitive to higher thresholds, than the SDPA and SCPA strategies, which are both able to serve 100\% of all parcels for all investigated $\tau_{th}$ for the \textit{Full RPP Integration}. Looking at Figure~\ref{fig:calib_served_persons} one can see the number of served customers for varying assignment thresholds. It shows that for all strategies, a notable decline in service rate for passengers cannot be observed. The SCPA and SDPA strategies are even able to serve additional passengers compared to the \textit{Status Quo}, which can presumably be traced back to increased vehicle availability within the network due to parcel pick-up and drop-off trips which can provide better coverage of the service area. Figure~\ref{fig:calib_vkm} displays the overall fleet kilometers traveled throughout the whole day. One can observe that the comprehensive (CDPA) strategy even produces a lower driven distance than the \textit{Status Quo} without the integrated parcel delivery. Compared to the fleet KM to serve only the parcels in the \textit{Status Quo} ($2,614$~km on average), this equals a reduction share of 48\% in traveled distance for $\tau_{th} = 0.8$. This means, that only $1,252$~km compared to the~\textit{Status Quo} were needed in the integrated transport approach. However, the slightly lower number of served customers and parcels compared to the \textit{Status Quo} have to be considered. The SCPA strategy and especially the SDPA approach, however, lead to considerably higher driven distances compared to the \textit{Status Quo}. Another interesting aspect in Figure~\ref{fig:calib_served_parcels} is that the subsequent strategies (SDPA and SCPA) can accommodate nearly all parcels with all thresholds for the \textit{Full RPP Integration}. This means that all parcels are picked-up because even with a high threshold the depots are very often located on the route of the vehicle fleet. Nevertheless, at the end of the day the collected but not yet dropped-off parcels, are then delivered, which also contributes to the significantly increased vehicle kilometers of the fleet using these strategies. Both, SDPA as well as SCPA, do not consider the destination of parcels when picking them up, which can be a big disadvantage when delivering them finally because parcel destinations might be distributed throughout the whole network.  For the \textit{Moderate RPP Integration} the SDPA and SCPA approaches produces less fleet distance with higher $\tau_{th}$ values, which corresponds to the falling numbers of served parcels. Figure~\ref{fig:calib_util} shows the fleet utilization, which of course leads to similar results as Figure~\ref{fig:calib_vkm}. One can observe that depending on the chosen scenario and assignment strategy the fleet utilization over the whole day varies between 54\% and 61\%.

The evaluations have shown that the \textit{Moderate RPP Integration} performs consistently worse than the \textit{Full RPP Integration}: Significantly fewer parcels can be served while the experience for passengers does not deteriorate much applying the \textit{Full RPP Integration}. Therefore, only full integration is considered for further analysis. Additionally, the threshold parameter will be fixed to $\tau_{th} = 0.8$, the value where a first notable decline in served parcels can be observed for the CDPA strategy.

\begin{figure}[!th]
    \centering
    \includegraphics[width=1.0\textwidth]{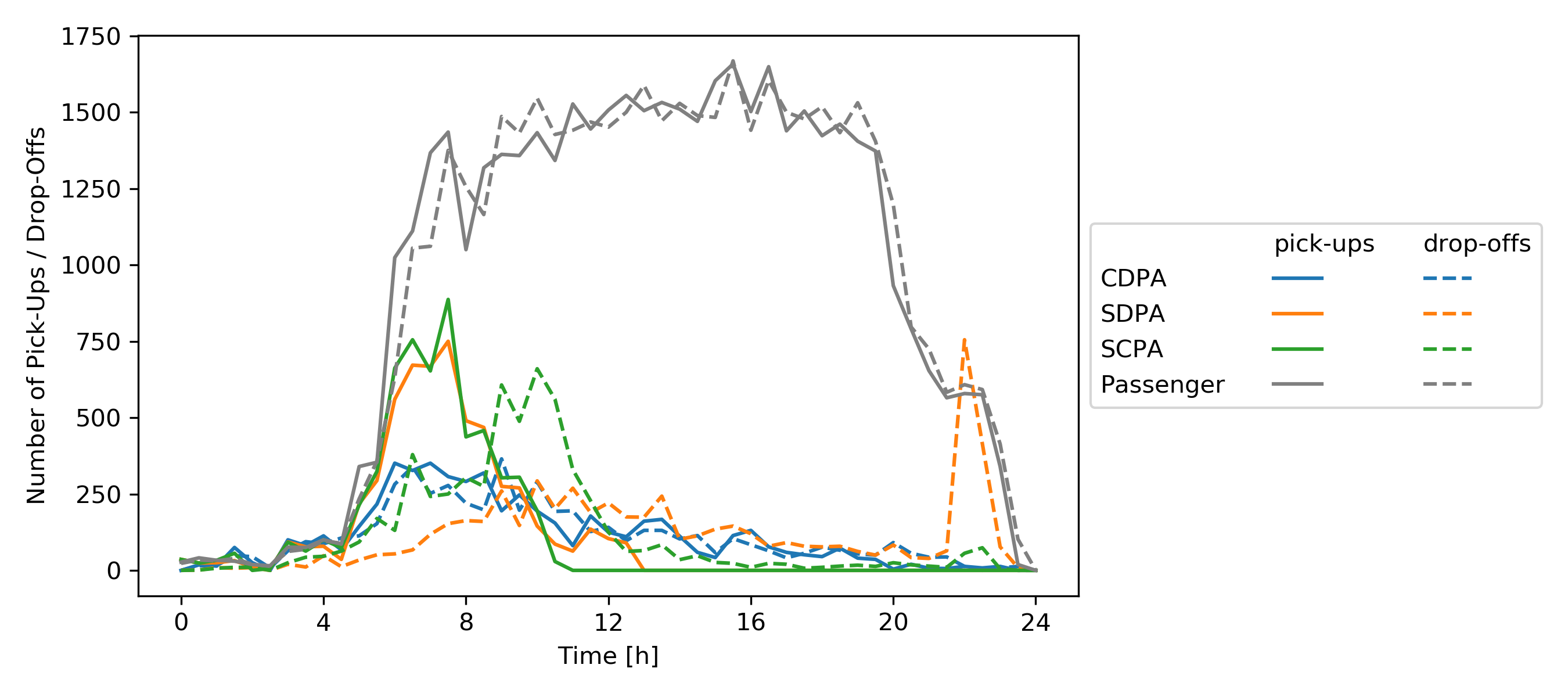}
    \caption{Time-dependent pick-ups and drop-offs of parcels for different parcel assignment strategies and threshold parameters $\tau_{th} = 0.8$. Full integration is considered in all scenarios shown.}
    \label{fig:pu_do}
\end{figure}

Figure~\ref{fig:pu_do} shows the temporal distribution of pick-ups and drop-offs for passengers and parcels for the different assignment strategies. Most passenger demand starts in the morning at around 6am and lasts until the late evening at around 8pm. It can be observed that parcel delivery only takes place, while there is passenger demand to exploit the already existing passenger trips for parcel delivery. Strategies using off-peak deliveries from other studies could easily be added on top, which would increase the capacity for parcel transport even further.

For the comprehensive (CDPA) strategy the temporal distribution for pick-ups and drop-offs of parcels is similar indicating a rather fast delivery after pick-up, resulting from the simultaneous assignment of pick-up and drop-off. Therefore, the vehicle is actively routed toward drop-off locations. The subsequent (SDPA and SCPA) strategies tend to collect the parcels right at the beginning of the day at around 7am. The peak in the morning indicates, that it seems to be easy to find routes including the logistics depots. The SCPA strategy shows a higher success rate in delivering parcels during the day compared to the SDPA strategy, as nearly all parcels could be served and the number of drop-offs strongly declines after a peak around 11am. The SDPA strategy, however, shows a strong peak in the number of drop-offs at 10pm, when the parcels which are still on board the vehicles and could not be delivered so far are driven to their destinations.

\begin{figure}
     \centering
     \begin{subfigure}[b]{0.49\textwidth}
         \caption{CDPA}
         \includegraphics[width=\textwidth]{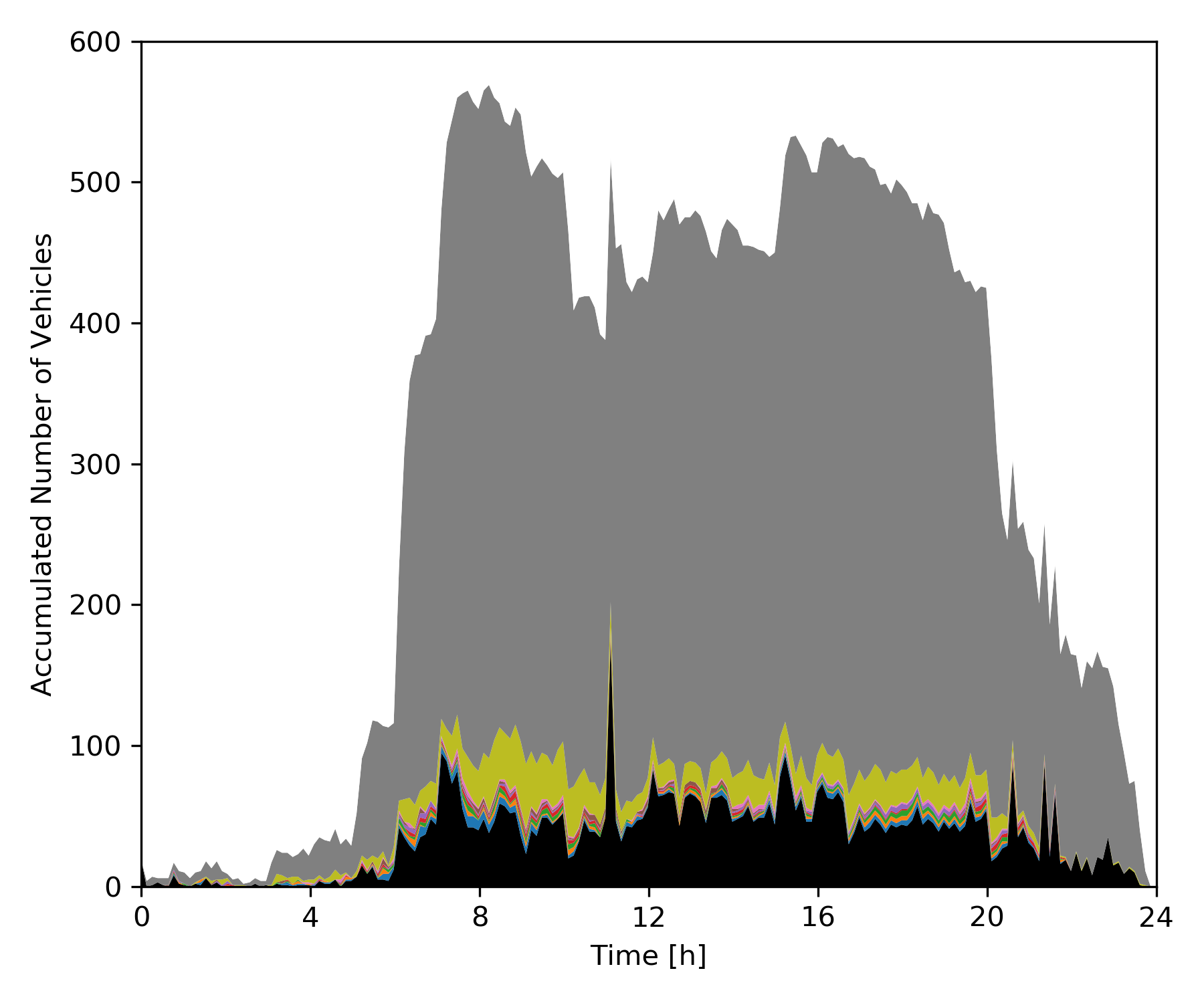}
         \label{fig:parcel_occ_states_CDPA}
     \end{subfigure}
     \hfill
     \begin{subfigure}[b]{0.49\textwidth}
         \caption{SDPA}
         \includegraphics[width=\textwidth]{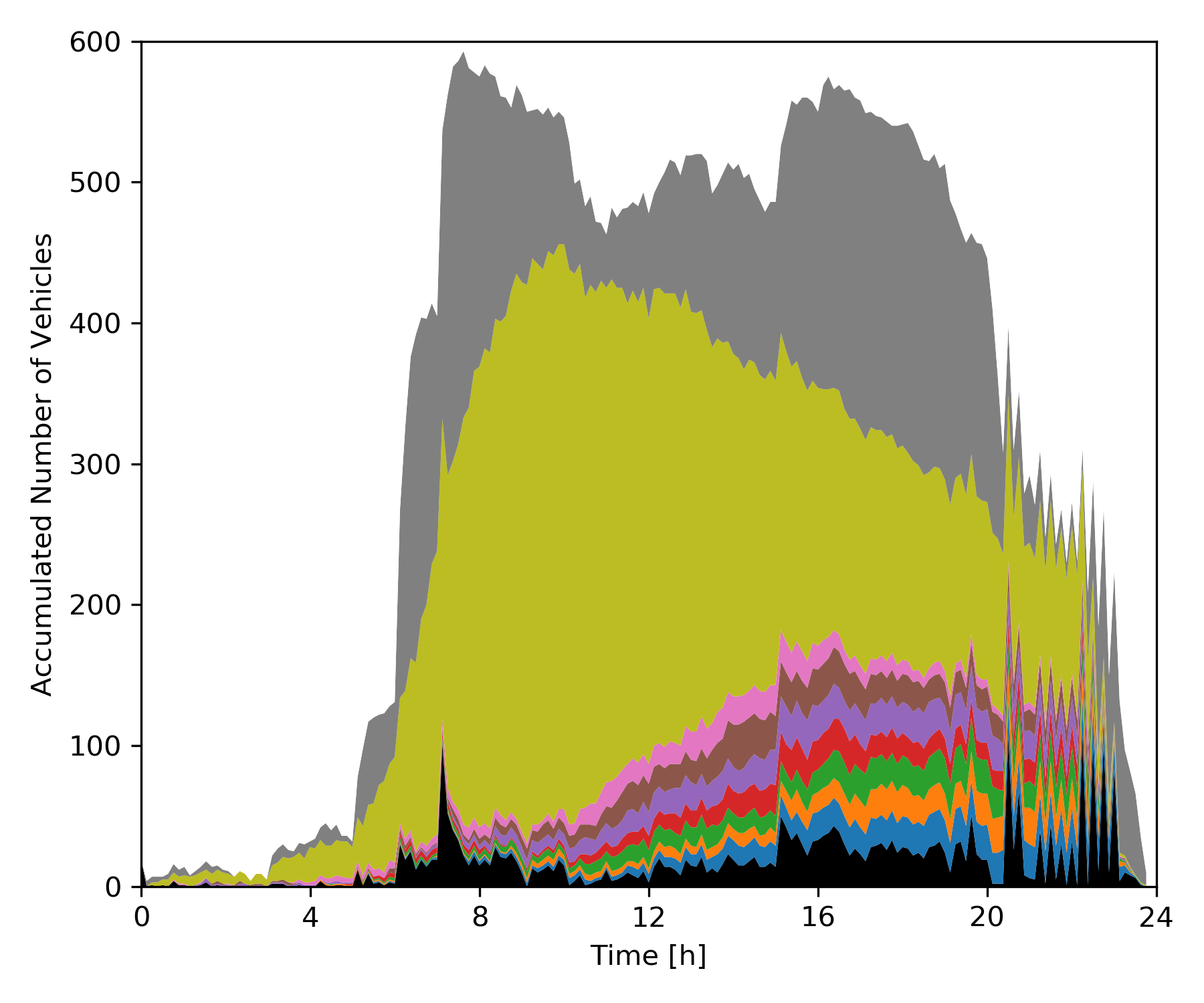}
         \label{fig:parcel_occ_states_SCPA}
     \end{subfigure}
     \hfill
     \begin{subfigure}[b]{0.49\textwidth}
         \caption{SCPA}
         \includegraphics[width=\textwidth]{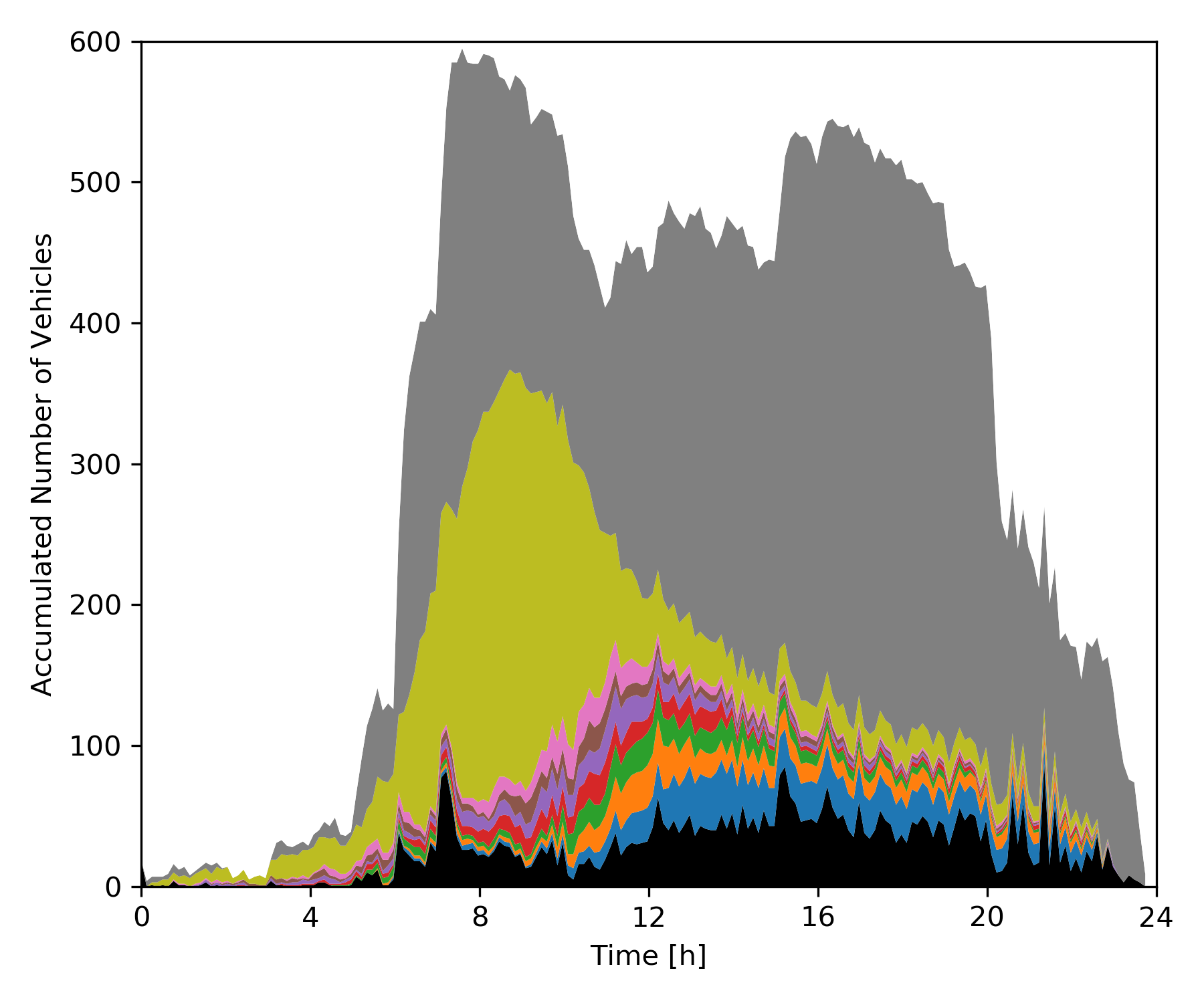}
         \label{fig:parcel_occ_states_SDPA}
     \end{subfigure}
     \hfill
     \begin{subfigure}[b]{0.49\textwidth}
         \caption{Legend}
         \includegraphics[width=\textwidth]{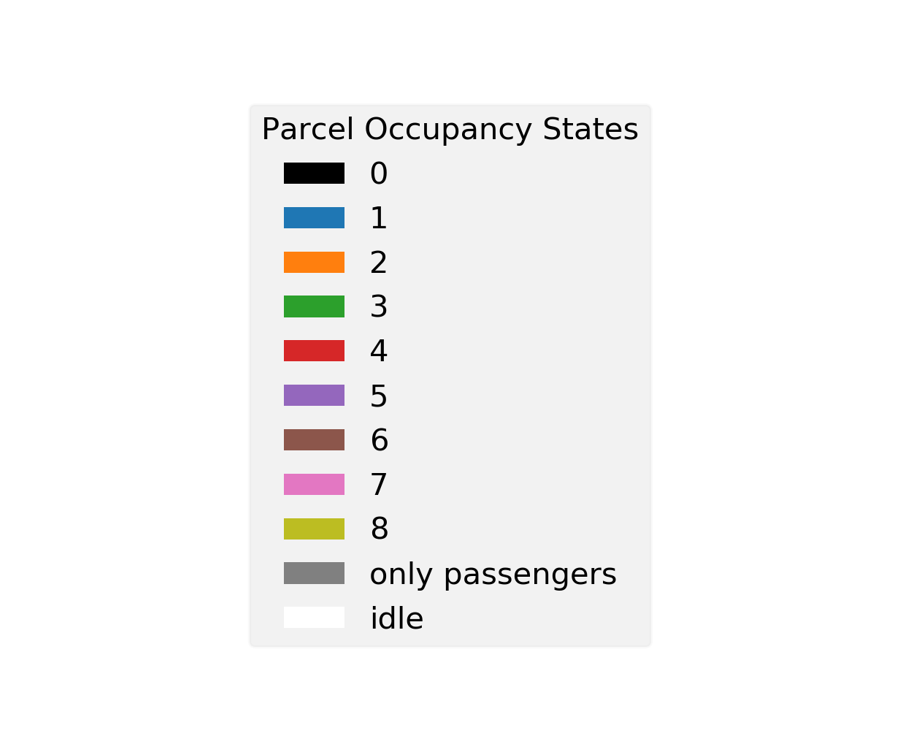}
         \label{fig:parcel_occ_states_legend}
     \end{subfigure}
        \caption{Parcel occupancy states of moving vehicles for different parcel assignment strategies. $\tau_{th} = 0.80$ and full integration is considered in all scenarios shown. (fleet size = 600)}
        \label{fig:parcel_occ_states}
\end{figure}

Figure~\ref{fig:parcel_occ_states} gives a detailed view of the temporal parcel occupancy states of the vehicles throughout the day. If no parcels or passengers are on board, vehicles are represented by a black color. White color is used for idle vehicles. It can be observed that for the CDPA strategy only around 50 of 600 vehicles have parcels on board during the day, indicating a rather fast delivery once a parcel is picked up, as stated previously. For the SDPA and SCPA strategies, most of the vehicles are filled with parcels in the morning, when they pass by the logistics depots and carry them around during the day. In the case of SCPA most of the parcels can be delivered during the day (Figure~\ref{fig:pu_do}), which results in low occupation states at the end of the day. Looking at the SDPA strategy, the occupation states at the end of the day are still high, which results in a delivery peak at around 10pm to drop-off the remaining parcels. In general, one can observe that the SCPA and SDPA strategies result in higher overall vehicle utilization, indicated by the area below the grey shape, compared to the CDPA strategy. This reflects the fact of higher fleet kilometers and the higher number of served passengers and parcels shown in Figure~\ref{fig:calib_th}.

\subsection{Variation of Logistics Demand}

\begin{figure}[!ht]
     \centering
     \begin{subfigure}[b]{0.49\textwidth}
         \caption{Served Parcels}
         \includegraphics[width=\textwidth]{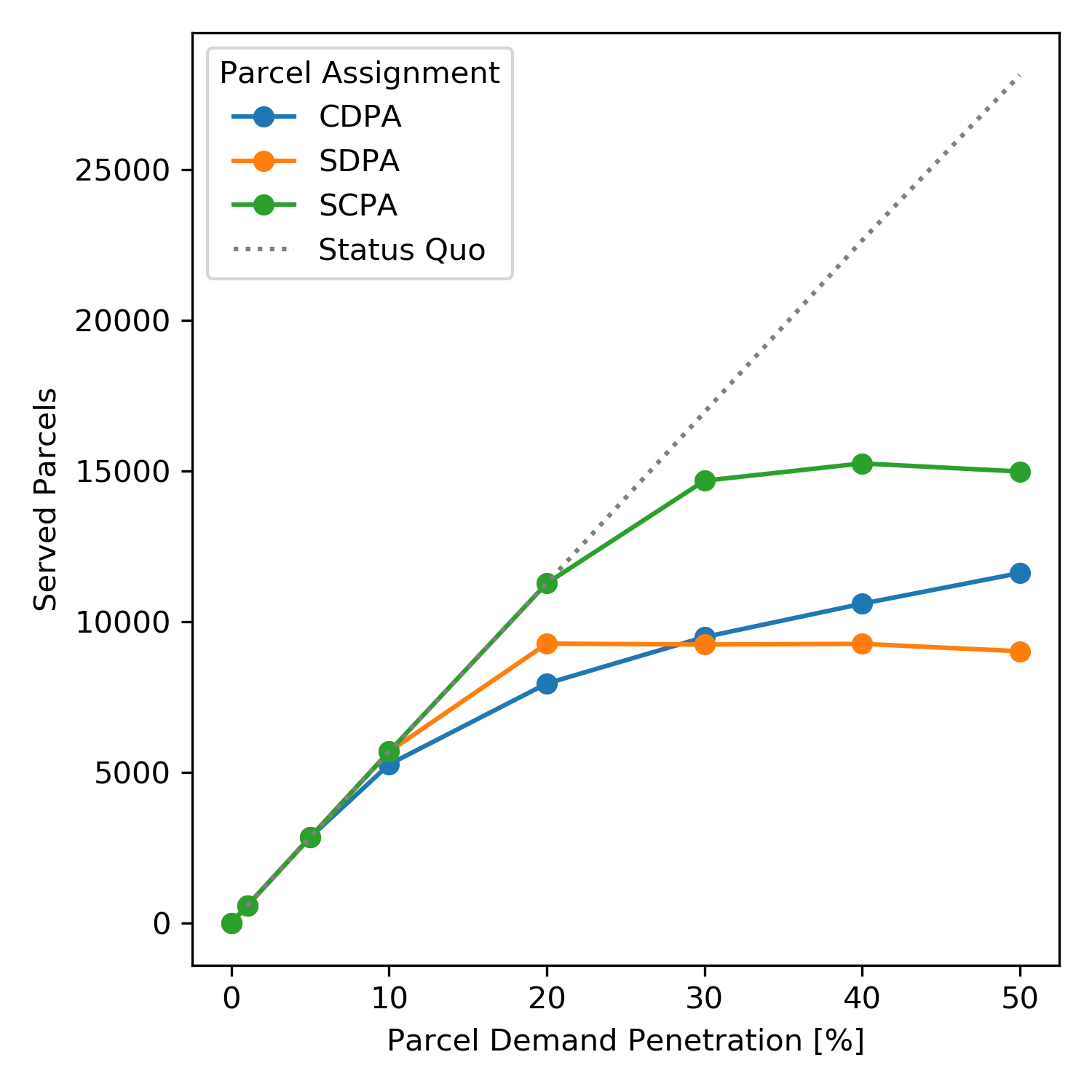}
         \label{fig:dem_served_parcels}
     \end{subfigure}
     \hfill
     \begin{subfigure}[b]{0.49\textwidth}
         \caption{Served Customers}
         \includegraphics[width=\textwidth]{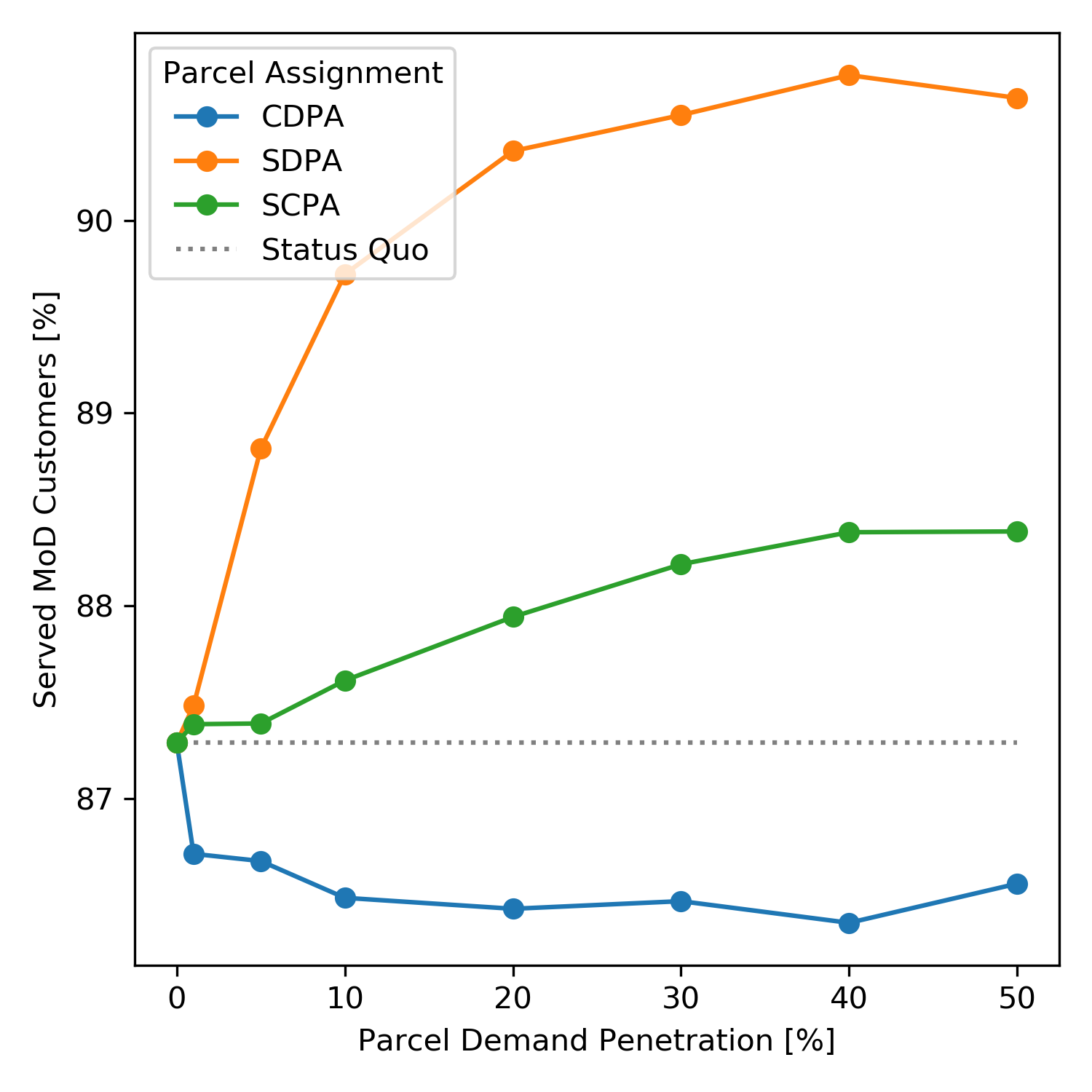}
         \label{fig:dem_served_persons}
     \end{subfigure}
     \hfill
     \begin{subfigure}[b]{0.49\textwidth}
         \caption{Fleet KM}
         \includegraphics[width=\textwidth]{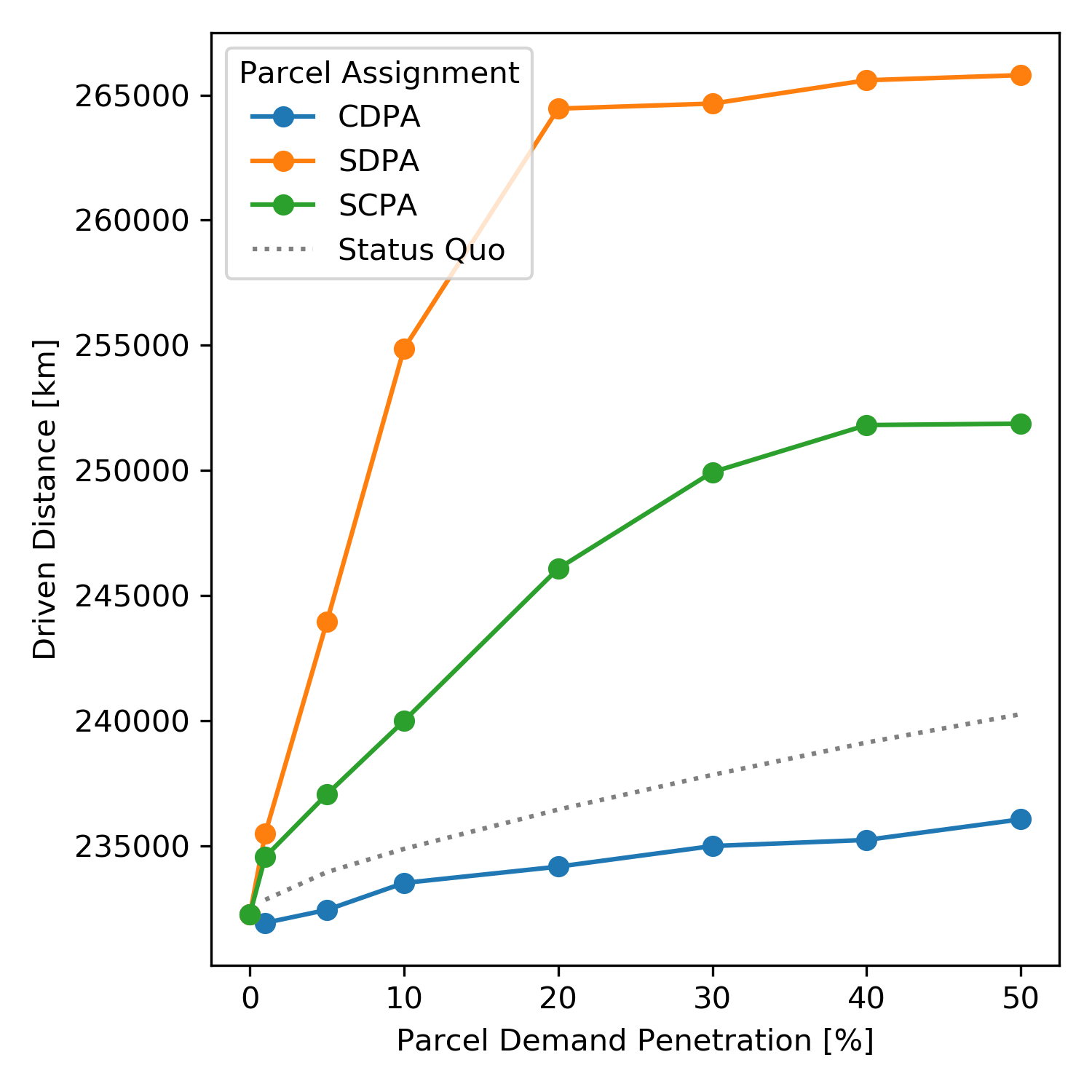}
         \label{fig:dem_vkm}
     \end{subfigure}
     \hfill
     \begin{subfigure}[b]{0.49\textwidth}
         \caption{Fleet KM Per Served Customer and Parcel}
         \includegraphics[width=\textwidth]{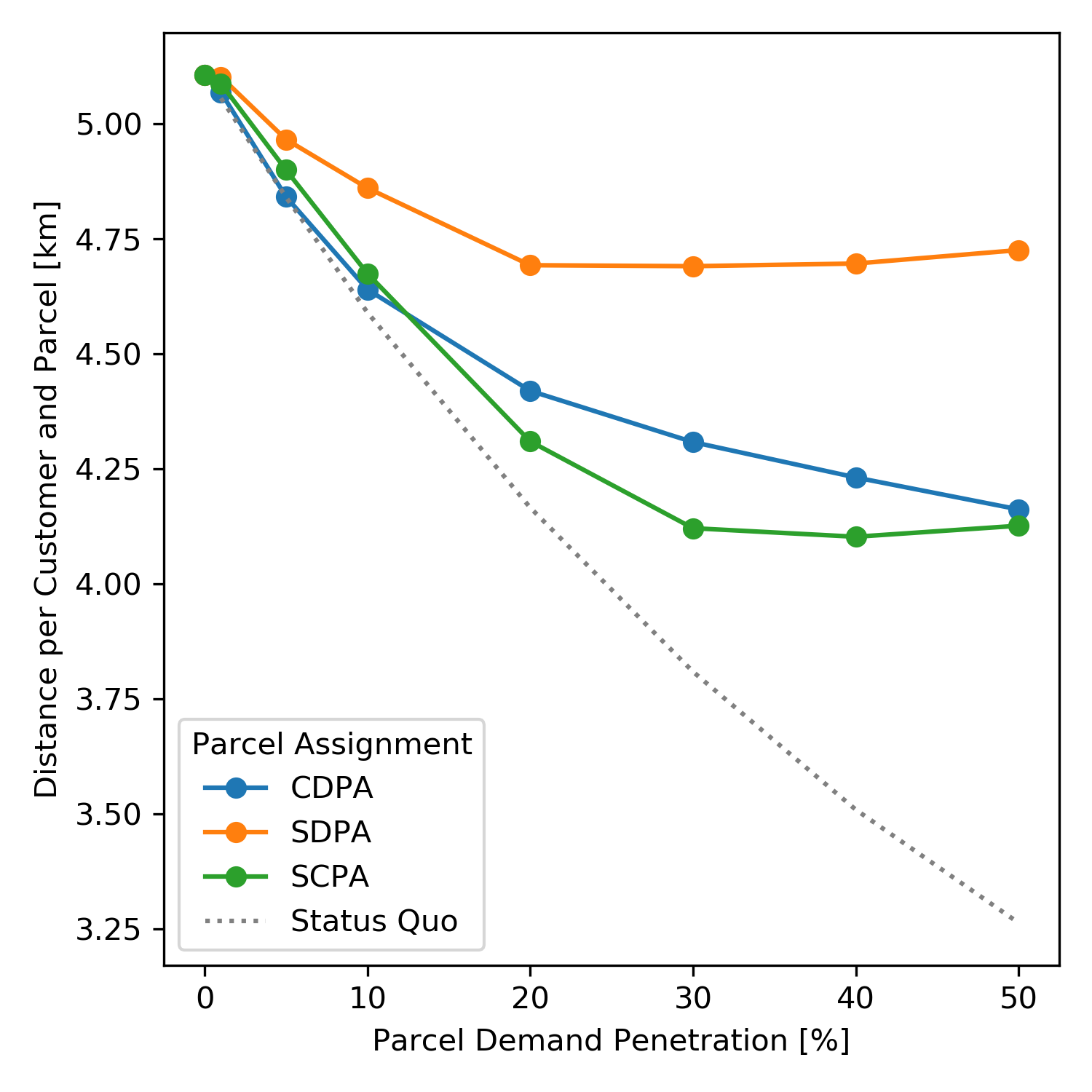}
         \label{fig:kmPerReq}
     \end{subfigure}
        \caption{Impact of varying parcel demand on the number of served parcels, served persons and fleet kilometers traveled. In all simulations $\tau_{th} = 0.8$ and full integration is considered.}
        \label{fig:dem}
\end{figure}

To investigate the system boundaries of the introduced RPP service, the logistics demand charged on the MoD fleet was varied. The overall data set of 56,000 parcel shipments is sub-sampled to shares of 1 to 50\%.

Figure~\ref{fig:dem_served_parcels} shows the total number of served parcels depending on the applied share of parcel demand. It becomes apparent that all strategies reach a certain limit of parcel transport. The CDPA and SDPA strategies match the \textit{Status Quo} until a parcel demand of approximately 10\%. The SCPA strategy is even able to serve a parcel demand of up to 20\% (~11200 parcels per day). As already observed, the SDPA and the SCPA approaches tend to pick up as many parcels as possible in the morning. Nevertheless, contrary to the SCPA strategy the SDPA strategy is not able to deliver the majority of parcels throughout the day, leading to a fleet state of close to full load, indicated by a horizontal plateau in Figure~\ref{fig:dem_served_parcels}. CDPA and SCPA on the other hand can also deliver most of the parcels during the service freeing up additional capacity to serve further parcels.

Figure \ref{fig:dem_served_persons} displays the relative number of served customers. Similar effects that have already been evaluated for the sensitivity of parameter $\tau_{th}$ in Figure~\ref{fig:calib_served_persons} can be observed: For any strategy, no notable drop in served customers can be observed, even for high parcel demand penetrations showing the potential of free capacity for a parcel service during the MoD operation. The higher fleet utilization seems to lead to a higher vehicle availability, resulting in an even slightly higher share of served customers (at least when using the assignment and re-balancing control strategies applied in this study).

When it comes to the traveled distance of the vehicle fleet in Figure~\ref{fig:dem_vkm}, it can be seen that only the CDPA strategy results in similar traveled distances compared to the \textit{Status Quo}. The CDPA strategy actually even stays below the \textit{Status Quo}, saving driving distance through the integrated transport of passenger and freight streams. The SCPA and SDPA tend to increase fleet kilometers heavily and produce around 15,000 km and 30,000 km more total distance when considering high parcel demand penetration rates. Until the number of served parcels stabilizes at around $20$\% penetration, Especially, SDPA shows a strong increase in the driven distance indicating a bad scaling behavior for finding effective routes for parcel pick-up and delivery. These findings are also in line with the insights and discussion presented in Figure~\ref{fig:calib_vkm}. However, similar scaling of fleet-driven distance with parcel penetration can be observed for the CDPA strategy. In this regime, a very low driven distance can be observed, especially for high parcel penetration rates (even lower than the \textit{Status Quo}). The down point of this strategy is that significantly fewer parcels are served in this regime. 
Looking at Figure~\ref{fig:kmPerReq} one can observe the fleet kilometers per served customer and parcel in relation to the parcel demand penetration. This quantity also takes the number of served customers and parcels into account when comparing fleet kilometers. Thereby, 'better' distance to request ratios can be achieved by either transporting more passengers or parcels or both, while keeping the driven distance low. All strategies show lower traveled distance per served request with rising parcel demand, indicating a better integration. However, with high parcel penetration efficient routes for parcel delivery can be found while all parcels are served in the \textit{Status Quo}, leading to the lowest values for fleet kilometers per served customer and parcel in this scenario. These results indicate that the modeled integration of parcel delivery into the MoD service is only reasonable for parcel penetration rates of around 10\%.

\section{Conclusion}

\subsection{Summary}

This paper investigates the integrated transport of passengers and freight, assuming a mobility-on-demand ride-pooling service, called Ride-Parcel-Pooling (RPP). Thereby, the operator of the vehicle fleet combines parcel delivery into the passenger routes of the MoD service. This study uses two different integration approaches, which are compared to the \textit{Status Quo} consisting of a separate logistics fleet and the MoD service serving passengers only. The \textit{Moderate RPP Integration} on the one hand, where parcels are only picked-up or dropped-off when no passengers are on board, is not very beneficial, as it results in lower numbers of served parcels and does not show better results in customer service quality compared to the full integration. The \textit{Full RPP Integration} scenario on the other hand, shows similar results in passenger waiting and detour times as the previous scenario and performs significantly better in the number of served parcels. The assignment of parcels is investigated using three different heuristic parcel assignment strategies (CDPA, SCPA and SDPA). Each of these strategies aims at inserting parcels into the schedules serving passengers with small detour for parcel pick-up and delivery. Because no explicit time constraints on parcel pick-up and delivery are employed, the assignment of parcel deliveries is separated from the parcel pick-ups in the SCPA and SDPA strategy contrary to the CDPA strategy. The results of a simulation case study for Munich, Germany suggest that the MoD service is not deteriorated by the integration of logistics services and is able to serve nearly all parcels until a parcel-to-passenger ratio of at least 1:10. The SCPA and SDPA strategies are able to transport more parcels than CDPA and can even achieve an increase of served passengers compared to the \textit{Status Quo}, however they result in significantly higher driven distance. The CDPA approach, however can decrease driven distances compared to the \textit{Status Quo}, where two vehicle fleets serve passenger and logistics demand independently.

\subsection{Discussion}

The presented RPP service offers a good opportunity to increase the utilization of MoD ride-pooling vehicle fleets. It could, assuming the existence of a MoD service, already today complement existing logistics services and extend the best practices in the sector. Thereby, it especially offers a solution to use cases, where conventional logistics approaches can not create any bundling effects, i.e. combining multiple parcels in one logistics vehicle. Furthermore, RPP has the potential to reduce the number of vehicles on the urban street network, by integrating passenger and logistics flows in one fleet of vehicles.

This paper examines a forward logistics use case, meaning that the parcels are transported from a depot to the recipient. In reality, this is the most relevant, but not the only logistics form. Reverse logistics from the sender to a consolidation center and courier services between two individuals could be interesting for future analysis. In the case of integrating courier services into the MoD ride-pooling service, a relatively cost-effective local distribution system could be built up. Especially, considering developments in autonomous driving, the integration into autonomous mobility-on-demand (AMoD) services could further decrease operational costs, by eliminating the need to pay a driver. This decrease in cost might further lead to decreased fares and therefore amplified demand, enhancing scaling properties of ride-pooling services that will also translate to more efficient options for parcel delivery integration. In the future, one could even imagine multi-purpose autonomous vehicles, which are able to convert passenger space into logistics space and vice versa. Such a vehicle concept could further promote the applicability of an integrated passenger and freight MoD service. A crucial aspect here is the time taken to load and unload the parcels. This study assumed that both loading and unloading are linked to a finely distributed parcel locker station system or that the senders and recipients take over or hand over the parcels at the vehicle, resulting in a relatively small boarding time of one minute. In reality, significantly higher times could arise, especially at the current stage without a high number of parcel lockers, which could deteriorate passenger satisfaction in particular. Overall, RPP might be a chance for local stores to compete with online shops, by offering fast delivery and a sustainable way of shopping. If the service can succeed in reducing the overall kilometers traveled in a city, by pooling passengers and parcels, this could have substantial benefits for the city and its residents. Not only it could induce a more livable city environment, by reducing the number of cars and traffic load in cities, but lead to a more sustainable way of city transportation without compromising on the customer experience, which MoD services offer.

\subsection{Future Work}

Generally, one can observe that the integration of a secondary parcel demand into the existing schemes of a MoD ride-pooling system is possible. The CDPA approach already offers promising results, although this heuristic approach is not of great complexity. The SCPA and SDPA strategies do still offer lots of room for improvement. In theory, at least a similar fleet performance should be achievable for a separated parcel pick-up and delivery assignment within the SCPA and SDPA strategies compared to the CDPA strategy. Several improvements can thereby be tested in the future for these strategies: Firstly, the decision of an assignment in the SCPA and SDPA strategies is made based on a threshold detour controlled by the parameter $\tau_{th}$ which is the same for the pick-up and the drop-off. Results showed that it seemed to be more probable to assign the pick-up compared to the delivery of a parcel. An asymmetric assignment threshold for pick-up and delivery could thus amplify efficiency. Secondly, the assignment of a parcel origin and therefore the delivery of a parcel by a certain vehicle does not consider the destination of parcels that are already assigned. It could be beneficial to assign parcels with similar destinations to the same vehicle. And thirdly, the delivery of parcels could be actively integrated into the re-balancing algorithm, i.e. if a lack of vehicle supply is recognized in a certain zone, vehicles with parcel deliveries in the corresponding zone could be prioritized for the corresponding re-balancing trip. Furthermore, future research could investigate the influences of a variation in loading and unloading times for parcels for finding the influence on customer waiting and travel times and by that customer satisfaction. Nevertheless, the overall goal should be the development of an integrated control algorithm that optimizes vehicle schedules and decides on parcel and passenger assignment in a unified optimization framework.

Apart from that, the presented RPP service could be extended by integrating vehicle idle times into the availability for logistics services. This could increase the system capacity for logistics services strongly and already developed efficient algorithms for planning routes for pick-up and delivery could be exploited. Nevertheless, care has to be taken on when and how many vehicles perform these pure logistics services because these vehicles will no longer be available for passenger transport. Additionally, a further increase in vehicle kilometers traveled is expected.

Looking at the RPP service definition, one could think of multiple extensions. The current service intentionally did not impose pickup and delivery time windows for parcels. However, this representation of parcels excludes certain types of logistics services that could also be use cases for RPP, for example, food and grocery, drugs and medicine, or high priority parcels. Furthermore, the effects of different logistics service forms (e.g. same-day vs. next-day delivery or forward vs. reverse vs. courier logistics) on the RPP performance could be investigated. It would be particularly interesting to look at edge cases that are currently financially not interesting for logistics companies due to low bundling effects and to investigate whether RPP could remedy this situation. Last but not least, constraints for the design of such a multi-purpose vehicle could be examined. Depending on the use case the optimal parcel capacity for a RPP vehicle could be determined in future studies. Additionally, a dynamic allocation of capacity based on the current demand to convert seats into parcel storage and vice-versa could be implemented. 

\section*{Conflict of Interest Statement}
The authors declare that they have no known competing financial interests, or personal relationships that could have appeared to influence the work reported in this paper.

\section*{Author Contributions}
Study conception and design: FF, RE, FD, KB, FB; data collection: FF, RE; analysis and interpretation of results: FF, RE, FD; simulation model: RE, FD, FF; draft manuscript preparation: FF, RE, FD, KB, FB. All authors reviewed the results and approved the final version of the manuscript.

\section*{Funding}
The German Federal Ministry of Transport and Digital Infrastructure provides funding through the project “TEMPUS” with grant number 01MM20008K. The authors remain responsible for all findings and opinions presented in the paper.

\printbibliography
\end{document}